\documentclass[a4paper,11pt]{article}
\usepackage{jcappub} 
\usepackage{amssymb}

\allowdisplaybreaks
\graphicspath{{./}{./figs/}{./figure/}}

\title{Testing $\Lambda$LTB as an alternative to dynamical dark energy}

\author[a,c,d]{Yukang Zhang}
\author[a,b,c,d]{and Pengjie Zhang}

\affiliation[a]{School of Physics and Astronomy,
Shanghai Jiao Tong University, Shanghai 200240, China}

\affiliation[b]{Tsung-Dao Lee Institute,
Shanghai Jiao Tong University, Shanghai 201210, China}

\affiliation[c]{State Key Laboratory of Dark Matter Physics,
Shanghai 200240, China}

\affiliation[d]{Key Laboratory for Particle Astrophysics and Cosmology (MOE)/
Shanghai Key Laboratory for Particle Physics and Cosmology,
Shanghai 200240, China}

\emailAdd{yukangzhang@sjtu.edu.cn}
\emailAdd{zhangpj@sjtu.edu.cn}

\abstract{
A cosmological constant $\Lambda$ in a universe weakly violating the cosmological principle (CP) can mimic a dynamical dark energy (DDE) in a FRW universe, and therefore challenging the recently reported DDE evidences. We test this scenario in the form of $\Lambda$LTB cosmology, against two data combinations. (1) DESI DR2 BAO, Planck CMB, together with any of the four SNe Ia compilations (UNION3, Pantheon+, DES SN5YR, and DES Dovekie) show no preference over DDE or $\Lambda$LTB. (2) The inclusion of the kinematic Sunyaev Zel'dovich (kSZ) effect rules out a large portion of the CP violation parameter space allowed by (1). However, the existence of a $\sim 300$ Mpc$/h$ region with $\sim 10\%$ matter density excess is still allowed by the current data, and therefore remains a viable alternative to DDE. To facilitate further tests, we present analytical expressions of the distance-redshift relation and the kSZ effect in a universe with weak CP violation. 
}

\begin{document}
\maketitle
\flushbottom

\section{Introduction}\label{sec:intro}

The $\Lambda$CDM cosmology, being the standard model of modern cosmology, is challenged by various cosmological observations. In particular, DESI baryon acoustic oscillation (BAO) measurements together with CMB, SNe Ia and weak gravitational lensing reveal deviations from the $\Lambda\mathrm{CDM}$ model, pointing toward dynamical dark energy (DDE) exhibiting Quintom behavior \cite{DESIDR1ResultsVI,DESIDR2ResultsII,DESIDR2DDE,DESCollaboration2026,feng2005dark} . If verified, this will be a groundbreaking discovery of fundamental physics. However, the DDE interpretation is obtained within the FRW framework, which assumes the cosmological principle (CP) that the universe at sufficiently large scale is homogeneous and isotropic \cite{Ishak2019}. CMB observations verify the isotropy of the universe viewed from us \cite{Planck2018VII}. However, radial inhomogeneities are more difficult to rule out, since they are entangled with temporal evolution along our past light cone \cite{Mustapha1997,Valkenburg2014}. Such large-scale radial inhomogeneities alter the distance-redshift relation and can in principle mimic the effects of DDE. Therefore, a crucial task is to distinguish the two scenarios, namely a DDE in a FRW universe versus a $\Lambda$ in a non-FRW universe. 

While DDE in a FRW universe has been extensively studied, the alternative possibility has received comparatively less attention in this context. To examine such a non-FRW scenario, we consider the Lemaître–Tolman–Bondi (LTB) spacetime \cite{lemaitre1931expanding,Tolman1934,Bondi1947} with a cosmological constant, a spherically symmetric but radially inhomogeneous solution. For an observer located near the symmetry center, the sky can remain nearly isotropic, consistent with cosmic microwave background (CMB) observations, while light-cone observables may still differ from their counterparts in the $\Lambda\mathrm{CDM}$ model.

Early LTB studies explored a more radical possibility: radial inhomogeneities could entirely account for the accelerated cosmic expansion without a cosmological constant \cite{Celerier2000,Zibin2008}. This scenario is now severely constrained because large radial inhomogeneities can induce large coherent bulk flows, and generate large kinetic Sunyaev–Zel'dovich (kSZ) signals \cite{GarciaBellido2008,Yoo2010,ZhangStebbins2011,Bull2012,Redlich2014}. A milder approach retains the cosmological constant while introducing weak radial inhomogeneities, leading to the $\Lambda\mathrm{LTB}$ model. Such models have been employed to describe low-redshift distance observables \cite{MarraPaakkonen2010,RomanoChen2014,Tokutake2018,HoscheitBarger2018,Ding2019}. Their consistency with kSZ constraints has also been investigated, and specific radial profiles remain viable \cite{HoscheitBarger2018,Ding2019}. Recent evidence for DDE has renewed interest in whether deviations from $\Lambda\mathrm{CDM}$ could instead arise from cosmic inhomogeneities \cite{Ginat2026} or local effects \cite{Gialamas2025}. In this context, Camarena et al. recently confronted a $\Lambda\mathrm{LTB}$ model with DESI-era cosmological observations \cite{Camarena2025}.

However, efficient calculations in $\Lambda\mathrm{LTB}$ models remain challenging. Although exact implicit solutions exist for the time evolution, evaluating the metric functions requires the numerical inversion of elliptic integrals at every spacetime point. Furthermore, predicting kSZ signals entails computationally demanding geodesic ray-tracing from off-center scattering locations. These requirements substantially hinder efficient exploration of the cosmological parameter space. Consequently, a fast, analytical framework capable of computing both observables is strongly needed.

In this work, we develop such a framework. Since the departures from $\Lambda\mathrm{CDM}$ reported by DESI are mild, implying a universe with weak CP violation if attributed to radial inhomogeneities, we thus treat the LTB metric as a first-order perturbation around a spatially flat FRW background. Substituting this perturbed metric into the Einstein field equations, we solve the linearized equations perturbatively to obtain explicit expressions for the metric functions. By solving the null geodesic equations along our past light cone, we express the spacetime coordinates as functions of redshift. Combining these light-cone trajectories with the metric-defined transverse and radial distances, we derive explicit analytic distance–redshift relations. To incorporate the kSZ information, we then model the kSZ angular power spectrum. Because kSZ predictions are most naturally formulated in the conformal Newtonian gauge, we map the perturbed LTB solution into this gauge via a first-order coordinate transformation, exploiting the fact that both describe linear perturbations around a flat FRW background. This allows us to bypass computationally expensive geodesic ray-tracing and directly derive the kSZ angular power spectrum using the standard remote-dipole formalism \cite{ZhangJohnson2015,Terrana2017}. The resulting framework enables efficient joint likelihood analyses combining SNe Ia compilations, BAO, and CMB data, both with and without kSZ constraints. 

Applying this framework to four SNe Ia datasets (each combined with BAO and CMB measurements), we find that the $\Lambda\mathrm{LTB}$ model yields fits broadly comparable to those of the DDE model, both before and after imposing the kSZ constraint. Although the kSZ constraint substantially restricts large radial inhomogeneities, a parameter region with a matter excess of $\sim 10\%$ and a characteristic scale of $\sim 300\,h^{-1}\mathrm{Mpc}$ remains viable. Thus, current observations do not uniquely require a dynamical dark energy interpretation of distance-redshift relation, leaving large-scale inhomogeneity as a viable alternative.

The paper is organized as follows. In Sec.~\ref{sec:methods}, we construct the first-order $\Lambda$LTB model, derive the distance--redshift relation, and obtain constraints from distance data alone. In Sec.~\ref{sec:ksz}, we transform the first-order solution to the conformal Newtonian gauge and compute the associated kSZ angular power spectrum. We then impose the upper-bound constraint $D_{\ell=3000}^{\rm kSZ}\leq1\,\mu{\rm K}^2$ and present the combined constraints. In Sec.~\ref{sec:discussion}, we discuss the implications of the results and summarize the limitations of the present framework.

\section{Distance-data Constraints on the Radial Inhomogeneity}
\label{sec:methods}

\subsection{First-order $\Lambda$LTB distance observables}

To derive the distance--redshift relation within the $\Lambda$LTB framework, we begin with the LTB metric:
\begin{equation}
ds^2=-dt^2+\frac{R'^2(t,r)}{1+2E(r)}\,dr^2+R^2(t,r)\,d\Omega^2 .
\label{eq:ltb_metric_distance}
\end{equation}
where primes and overdots denote derivatives with respect to $r$ and $t$, respectively.

Substituting this metric into the Einstein field equations yields the Friedmann-like equation
\begin{equation}
\dot R^2(t,r)=\frac{2M(r)}{R(t,r)}+2E(r)+\frac{\Lambda}{3}R^2(t,r).
\label{eq:ltb_friedmann_distance}
\end{equation}
We model the LTB spacetime as a weak radial perturbation about a spatially flat FRW background
\begin{equation}
R(t,r)=a(t)r[1+B(t,r)],\qquad |B|\ll1,
\qquad
E(r)=E_1(r),\qquad |E_1|\ll1 .
\label{eq:ltb_ansatz_distance}
\end{equation}
Here, $a(t)$ is the background scale factor, $B(t,r)$ is the perturbation to the areal radius, and $E_1(r)$ is the spatial curvature perturbation relative to the flat background. We parameterize the mass function as
\begin{equation}
M(r)=M_0r^3[1+m(r)],
\qquad
M_0=\frac12H_0^2\Omega_{m0},
\label{eq:mass_function_distance}
\end{equation}
where $m(r)$ represents the enclosed present-day density perturbation
\begin{equation}
m(r)=\frac{3}{r^3}\int_0^r\delta_0(x)x^2dx,
\qquad
\rho_0(r)=\bar\rho_0[1+\delta_0(r)] .
\label{eq:enclosed_delta_distance}
\end{equation}

By linearizing Eq.~\eqref{eq:ltb_friedmann_distance} with respect to $B$, $E_1$, and $m$, we separate the system into the background Friedmann equation and a first-order differential equation governing radial deviation from the background. The zeroth-order component corresponds to a spatially flat $\Lambda$CDM background
\begin{equation}
H^2(a)=H_0^2\left(\Omega_{m0}a^{-3}+\Omega_{\Lambda0}\right),
\qquad
\Omega_{\Lambda0}=1-\Omega_{m0},
\qquad
H(a)\equiv\frac{\dot a}{a}.
\label{eq:bg_H_distance}
\end{equation}

The first-order perturbation satisfies
\begin{equation}
2H\dot B+3H^2\Omega_m(a)B
=H^2\Omega_m(a)m(r)+\frac{2E_1(r)}{a^2r^2},
\qquad
\Omega_m(a)=
\frac{\Omega_{m0}a^{-3}}
{\Omega_{m0}a^{-3}+\Omega_{\Lambda0}} .
\label{eq:B_eom_distance}
\end{equation}
The corresponding solution is
\begin{equation}
B(a,r)=H(a)
\left[
C_{\rm dec}(r)
+\frac{m(r)}{2}I_m(a)
+\frac{E_1(r)}{r^2}I_E(a)
\right],
\label{eq:B_general_distance}
\end{equation}
where $C_{\rm dec}$ is the decaying mode and
\begin{equation}
I_m(a)=\int_0^a
\frac{\Omega_m(\tilde a)}{\tilde a H(\tilde a)}\,d\tilde a,
\qquad
I_E(a)=\int_0^a
\frac{d\tilde a}{\tilde a^3H^3(\tilde a)} .
\label{eq:Im_IE_distance}
\end{equation}
We impose a homogeneous big bang time so $C_{dec}=0$, and fix the radial gauge by imposing
\begin{equation}
R(t_0,r)=r .
\label{eq:present_gauge_distance}
\end{equation}
These conditions remove the decaying mode and fix the curvature perturbation
\begin{equation}
\frac{E_1(r)}{r^2}
=
-\frac{m(r)}{2}
\frac{I_m(1)}{I_E(1)} .
\label{eq:E1_fixed_distance}
\end{equation}
The areal-radius perturbation thus simplifies to
\begin{equation}
B(a,r)=\frac{m(r)}{2}H(a)F(a),
\qquad
F(a)=I_m(a)-\frac{I_m(1)}{I_E(1)}I_E(a).
\label{eq:B_solution_distance}
\end{equation}
Once the present-day density profile $m(r)$ is specified, Eqs.~\eqref{eq:E1_fixed_distance} and \eqref{eq:B_solution_distance} completely determine the first-order LTB geometry.

The transverse and radial distance observables are obtained by evaluating the metric along the past light cone. For a central observer,
\begin{equation}
D_M(z)=(1+z)R[t(z),r(z)],
\qquad
D_H(z)=\frac{R'}{\dot R'}[t(z),r(z)]
\label{eq:DM_DH_exact_distance}
\end{equation}
in units with $c=1$. To compute the light-cone coordinates perturbatively, we rewrite Eq.~\eqref{eq:ltb_metric_distance} in conformal time, $dt=a\,d\eta$, as
\begin{equation}
ds^2=a^2(\eta)
\left[
-d\eta^2+(1+2\mathcal C)\,dr^2
+r^2(1+2B)\,d\Omega^2
\right],
\label{eq:conformal_ltb_distance}
\end{equation}
with
\begin{equation}
\mathcal C(t,r)\equiv B(t,r)+r\,\partial_rB(t,r)-E_1(r).
\label{eq:C_def_distance}
\end{equation}
Incoming radial null geodesics satisfy
\begin{equation}
\frac{dr}{d\eta}\simeq-(1-\mathcal C),
\qquad
\frac{dz}{d\eta}=-(1+z)(\mathcal H+\partial_\eta\mathcal C),
\label{eq:null_geodesic_distance}
\end{equation}
where $\mathcal H \equiv aH$ is the conformal Hubble parameter. We expand the light-cone coordinates perturbatively as
\begin{equation}
\eta(z)=\eta^{(0)}(z)+\delta\eta(z),
\qquad
r(z)=r^{(0)}(z)+\delta r(z),
\qquad
r^{(0)}(z)=-\eta^{(0)}(z)=\int_0^z\frac{d\tilde z}{H(\tilde z)} .
\label{eq:lightcone_expansion_distance}
\end{equation}
The scale factor entering light-cone quantities is evaluated at this perturbed conformal time. To first order,
\begin{equation}
a(\eta(z))
=
a\!\left(\eta^{(0)}(z)+\delta\eta(z)\right)
=
\frac{1}{1+z}
+
\left.\frac{da}{d\eta}\right|_{\eta^{(0)}(z)}
\delta\eta(z)
=
\frac{1}{1+z}
\left[
1+\frac{H(z)}{1+z}\delta\eta(z)
\right],
\label{eq:a_lightcone_distance}
\end{equation}
where we have used the background relation $a(\eta^{(0)}(z))=(1+z)^{-1}$.

With the boundary conditions $\delta\eta(0)=\delta r(0)=0$, the first-order corrections are given by
\begin{equation}
\delta\eta(z)=
\frac{1+z}{H(z)}
\int_0^z
\frac{[\partial_\eta\mathcal C]_{\rm lc}(\tilde z)}
{H(\tilde z)}\,d\tilde z,
\label{eq:deta_distance}
\end{equation}
\begin{equation}
\delta r(z)=
-\delta\eta(z)
-\int_0^z
\frac{[\mathcal C]_{\rm lc}(\tilde z)}
{H(\tilde z)}\,d\tilde z,
\label{eq:dr_distance}
\end{equation}
where $[X]_{\rm lc}$ denotes evaluation on the background light cone. Combining these relations with $R(t,r)=a(t)r[1+B(t,r)]$ gives
\begin{equation}
D_M(z)=
r^{(0)}(z)+[r^{(0)}B]_{\rm lc}
-\int_0^z
\frac{[\mathcal C]_{\rm lc}(\tilde z)}
{H(\tilde z)}\,d\tilde z-
\left[
1-\frac{H(z)r^{(0)}(z)}{1+z}
\right]\delta\eta(z),
\label{eq:DM_distance}
\end{equation}
and
\begin{equation}
D_H(z)=
\frac{1}{H(z)}
\left[
1+H_{,z}(z)\delta\eta(z)
-\frac{1+z}{H(z)}
[\partial_\eta\mathcal C]_{\rm lc}(z)
\right].
\label{eq:DH_distance}
\end{equation}
These expressions constitute the key distance-sector observables used below. Throughout this derivation, we adopt geometrized units ($G=c=1$). For numerical calculations, we restore physical units, reporting distances in Mpc and expansion rates in ${\rm km\,s^{-1}\,Mpc^{-1}}$.

The distances $D_M$ and $D_H$ derived above are geometric quantities. Their comparison with BAO measurements additionally requires accounting for the evolution of the standard ruler. Assuming a spatially uniform and isotropic physical BAO scale $r_d/(1+z_d)$ at the drag epoch $t_d$, followed by comoving evolution with the matter flow, we obtain the effective BAO ratios in the local-ruler approximation:
\begin{equation}
\left(\frac{D_M}{r_d}\right)_{\rm BAO}
=
F_M(z)\frac{D_M(z)}{r_d},
\qquad
\left(\frac{D_H}{r_d}\right)_{\rm BAO}
=
F_H(z)\frac{D_H(z)}{r_d}.
\label{eq:bao_corrected_ratios}
\end{equation}
These ratios correspond to the inverse angular and redshift separations of the BAO feature, respectively. Here, $z_d$ is the background drag-epoch redshift, and the ruler-evolution factors are \cite{MarraPaakkonen2010}
\begin{equation}
F_M(z)
=
\frac{1+z_d}{1+z}
\frac{R[t_d,r(z)]}{R[t(z),r(z)]},
\qquad
F_H(z)
=
\frac{1+z_d}{1+z}
\frac{R'[t_d,r(z)]}{R'[t(z),r(z)]}.
\label{eq:bao_ruler_factors}
\end{equation}
Expanding consistently to first order gives
\begin{equation}
F_M(z)
\simeq
1-\frac{H(z)}{1+z}\delta\eta(z)
+B[t_d,r^{(0)}(z)]-[B]_{\rm lc}(z),
\label{eq:bao_FM_linear}
\end{equation}
\begin{equation}
F_H(z)
\simeq
1-\frac{H(z)}{1+z}\delta\eta(z)
+\mathcal C[t_d,r^{(0)}(z)]-[\mathcal C]_{\rm lc}(z).
\label{eq:bao_FH_linear}
\end{equation}
The time-independent curvature term $E_1(r)$ cancels in the difference of $\mathcal C$ between the two epochs.

\subsection{Distance data and likelihood}

We denote the parameter set by $\Theta$, split into a homogeneous background sector and a radial inhomogeneity sector
\begin{equation}
\Theta
\equiv
\underbrace{\{H_0,\omega_b,\omega_c\}}_{\text{FRW background}}
\cup
\underbrace{\{\text{$\Lambda$LTB density-profile parameters}\}}_{\text{LTB density perturbation on the background}} .
\label{eq:theta_split_distance}
\end{equation}
In the baseline implementation, the sampled parameter vector is
\begin{equation}
\Theta_X=
\{H_0,\omega_b,\omega_c,m_{0,\rm amp},r_0,\mathcal M_X\},
\label{eq:theta_sampled_distance}
\end{equation}
where $\omega_b=\Omega_{b0}h^2$, $\omega_c=\Omega_{c0}h^2$, and $h=H_0/(100\,{\rm km\,s^{-1}\,Mpc^{-1}})$.
For the $\Lambda$LTB density profile, we adopt a Garcia-Bellido \& Haugbølle (GBH)-motivated profile parametrization \cite{GarciaBellido2008,Kenworthy2019}
\begin{equation}
m(r)=m_{0,\rm amp}
\frac{
1-\tanh[(r-r_0)/(2\Delta r)]
}{
1+\tanh[r_0/(2\Delta r)]
}.
\label{eq:gbh_profile_distance}
\end{equation}
For brevity, our baseline analysis fixes the transition-width ratio to $\Delta r / r_0 = 0.4$. Given a parameter vector, this profile uniquely determines $D_M(z)$ and $D_H(z)$ via Eqs.~\eqref{eq:B_solution_distance}, \eqref{eq:DM_distance}, and \eqref{eq:DH_distance}, enabling direct comparison with observational distance data.
Here, $m(r)$ denotes the enclosed present-day density perturbation, with the corresponding local density contrast given by $\delta_0(r) = m(r) + r m'(r)/3$. For the normalized profile in Eq.\eqref{eq:gbh_profile_distance}, $m(0) = \delta_0(0) = m_{0,\mathrm{amp}}$.

We use the DESI DR2 BAO data, incorporating both the published compressed distance vector and its covariance matrix \cite{DESIDR2ResultsII}. We analyze four Type Ia supernova compilations separately: Pantheon+ \cite{Scolnic:2021amr,Brout:2022}, DES SN5YR \cite{DES:2024tys}, DES Dovekie \cite{DESDovekie2025}, and Union3 \cite{Rubin:2023}. For each SNe Ia sample, we utilize the published Hubble-diagram data vector alongside its full statistical-plus-systematic covariance matrix. The theoretical distance modulus for a given SNe Ia sample $X$ is
\begin{equation}
\mu_{\rm th}(z)=
5\log_{10}
\left[
\frac{D_L(z)}{\rm Mpc}
\right]
+\mathcal M_X,
\qquad
D_L(z)=(1+z)D_M(z).
\label{eq:mu_distance}
\end{equation}
where the nuisance parameter $\mathcal{M}_X$ absorbs the absolute magnitude calibration of compilation $X$ and is marginalized over during inference. For Pantheon+, we restrict the sample to $z > 0.01$ to reduce contamination from peculiar-velocity dominated objects.

Compressed CMB constraints are implemented via the Planck Gaussian prior detailed in Appendix A of DESI DR2 Results II \cite{DESIDR2ResultsII}. Defining $\omega_{bc} \equiv \omega_b + \omega_c$, the data and model vectors are
\begin{equation}
\mathbf d_{\rm CMB}
=
\left(
\theta_\ast,\omega_b,\omega_{bc}
\right)^T
=
\begin{pmatrix}
0.01041\\
0.02223\\
0.14208
\end{pmatrix},
\qquad
\mathbf m_{\rm CMB}(\Theta)=
\left(
\theta_\ast^{\rm th},\omega_b,\omega_b+\omega_c
\right)^T,
\label{eq:cmb_data_distance}
\end{equation}
with
\begin{equation}
\theta_\ast^{\rm th}=\frac{r_s(z_\ast)}{D_M(z_\ast)}
\label{eq:theta_star_distance}
\end{equation}
and covariance
\begin{equation}
\mathbf C_{\rm CMB}=10^{-9}
\begin{pmatrix}
0.006621 & 0.12444 & -1.1929\\
0.12444 & 21.344 & -94.001\\
-1.1929 & -94.001 & 1488.4
\end{pmatrix}.
\label{eq:cmb_cov_distance}
\end{equation}
Early-Universe quantities are computed using \texttt{CAMB} \cite{LewisChallinorLasenby2000}. Standard pre-recombination physics is assumed to derive $z_*$, $z_d$, $r_s(z_*)$, and $r_d = r_s(z_d)$, under the fixed settings
\begin{equation}
T_{\rm CMB}=2.7255\,{\rm K},
\quad
N_{\rm eff}=3.046,
\quad
\sum m_\nu=0.06\,{\rm eV},
\quad
\Omega_k=0,
\quad
\tau=0.054 .
\label{eq:camb_settings_distance}
\end{equation}
Because late-time radial inhomogeneities are negligible at recombination and the drag epoch, the early-Universe calibrations of $r_s(z_*)$ and $r_d$ remain unaffected.

All likelihoods are assumed to be Gaussian
\begin{equation}
\chi_i^2=
(\mathbf d_i-\mathbf m_i)^T
\mathbf C_i^{-1}
(\mathbf d_i-\mathbf m_i),
\qquad
i\in\{{\rm BAO,SNe~Ia,CMB}\}.
\label{eq:chi2_distance}
\end{equation}
For each SNe Ia compilation, the joint likelihood is expressed as
\begin{equation}
\ln\mathcal L
=
-\frac12
\left(
\chi_{\rm BAO}^2+\chi_{\rm SNe~Ia}^2+\chi_{\rm CMB}^2
\right)
+{\rm const}.
\label{eq:joint_likelihood_distance}
\end{equation}
The priors used in the distance-sector analysis are summarized in Table~\ref{tab:distance_priors}. We perform parameter estimation using the affine-invariant ensemble sampler \texttt{emcee} \cite{ForemanMackey2013}, and evaluate Bayesian evidences via bridge sampling. Model comparisons are evaluated relative to two spatially flat baseline cosmologies: $\Lambda\mathrm{CDM}$ and $w_0w_a\mathrm{CDM}$. We quantify relative performance using $\Delta{\rm AIC}_{\Lambda{\rm CDM}}$ and the log Bayes factor $\ln K_{\Lambda{\rm LTB},\Lambda{\rm CDM}}$ for the former, and $\Delta\chi^2_{w_0w_a{\rm CDM}}$ for the latter.

\begin{table}[htbp]
\centering
\begin{tabular}{ll}
\hline\hline
Parameter & Prior \\
\hline
$H_0\,[{\rm km\,s^{-1}\,Mpc^{-1}}]$ & $U(50,90)$ \\
$\omega_b$ & $U(0.005,0.04)$ \\
$\omega_c$ & $U(0.001,0.30)$ \\
$\mathcal M_X$ & $U(15.0,35.0)$ \\
$m_{0,\rm amp}$ & $U(-0.2,0.5)$ \\
$r_0\,[{\rm Mpc}]$ & $U(30,3000)$ \\
\hline
\end{tabular}
\caption{Parameter priors for the $\Lambda\text{LTB}$ model in the distance-sector analysis. Here, $U(a,b)$ denotes a flat uniform prior bounded between $a$ and $b$. The parameter set includes cosmological background quantities ($H_0$, $\omega_b$, $\omega_c$), a distance calibration nuisance parameter ($\mathcal{M}_X$), and $\Lambda\mathrm{LTB}$ density profile parameters (amplitude $m_{0,\mathrm{amp}}$ and scale radius $r_0$).}
\label{tab:distance_priors}
\end{table}

\subsection{Constraints from SNe~Ia, BAO, and CMB data}

\begin{figure}[htbp]
\centering
\includegraphics[width=0.75\textwidth]{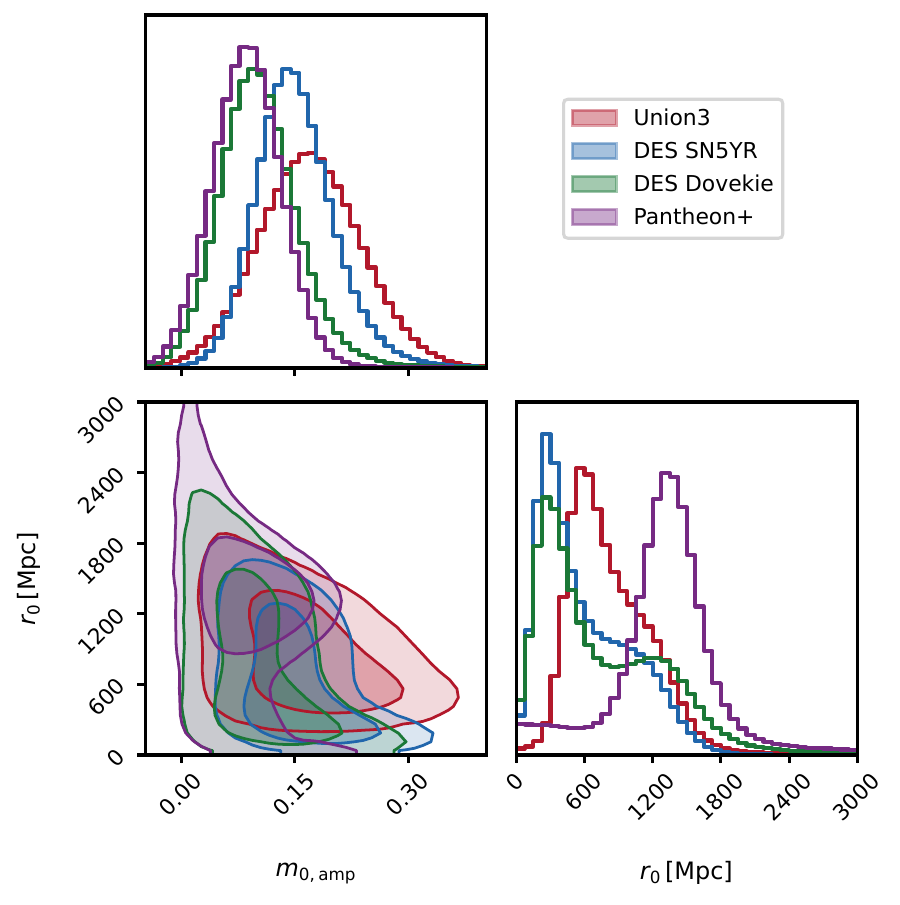}
\caption{
Marginalized 1D and 2D posterior distributions (68\% and 95\% C.L.) of $(m_{0,\mathrm{amp}}, r_0)$ under the baseline GBH-motivated profile. Results are presented for joint analyses combining DESI DR2 BAO and compressed CMB data with each of four SNe~Ia compilations: Union3, DES-SN5YR, DES Dovekie, and Pantheon+. All four combinations consistently favor a positive amplitude ($m_{0,\mathrm{amp}} > 0$), suggesting that the observer resides in a shallow overdense region, while the inferred characteristic scale $r_0$ exhibits substantial scatter across the SNe~Ia samples.
}
\label{fig:corner_width04}
\end{figure}

We present marginalized posterior constraints for the four joint analyses, each combining an individual SN~Ia compilation with DESI DR2 BAO data and the compressed CMB information. Fig.~\ref{fig:corner_width04} illustrates the constraints in the $(m_{0,\mathrm{amp}}, r_0)$ parameter subspace. The preferred profile amplitudes remain consistently positive across all four cases, with $m_{0,\mathrm{amp}}$ ranging from $0.087$ (Pantheon+) to $0.170$ (Union3). Similarly, the preferred characteristic scales span $r_0 \simeq 490\,\mathrm{Mpc}$ (DES SN5YR) to $r_0 \simeq 1320\,\mathrm{Mpc}$ (Pantheon+). Under this parametrization, a positive amplitude indicates that the distance data favor an observer situated within a shallow overdense region.

The parameter constraints and model-comparison metrics are summarized in Table~\ref{tab:distance_constraints}. Relative to spatially flat $\Lambda$CDM, the DES SN5YR combination gives strong support for the $\Lambda$LTB model, while Union3 gives moderate support. DES Dovekie and Pantheon+ give weaker differences, indicating comparable support for $\Lambda$LTB and $\Lambda$CDM. A direct comparison with spatially flat $w_0w_a\mathrm{CDM}$ demonstrates that both models fit the distance data to a similar degree. Because $\Lambda\mathrm{LTB}$ and $w_0w_a\mathrm{CDM}$ share the same number of parameters in this comparison, their small differences in best-fit $\chi^2$ indicate that weak radial inhomogeneity and dynamical dark energy describe current distance data equally well.

\begin{table}[htbp]
\centering
\renewcommand{\arraystretch}{1.15}
\begin{tabular}{lccccc}
\hline\hline
\shortstack{SNe~Ia compilation\\(with BAO+CMB)}
& $m_{0,\rm amp}$
& $r_0$ (Mpc)
& $\Delta{\rm AIC}_{\Lambda{\rm CDM}}$
& \shortstack{Log Bayes factor\\$\ln K_{\Lambda{\rm LTB},\Lambda{\rm CDM}}$}
& $\Delta\chi^2_{w_0w_a{\rm CDM}}$ \\
\hline
Union3
& $0.170^{+0.068}_{-0.066}$
& $760^{+440}_{-250}$
& $-6.89$
& $2.44$
& $1.22$ \\
DES SN5YR
& $0.148^{+0.052}_{-0.045}$
& $490^{+580}_{-260}$
& $-11.55$
& $4.24$
& $0.11$ \\
DES Dovekie
& $0.100^{+0.051}_{-0.045}$
& $590^{+750}_{-350}$
& $-5.14$
& $1.30$
& $-0.27$ \\
Pantheon+
& $0.087^{+0.043}_{-0.044}$
& $1320^{+290}_{-360}$
& $-3.90$
& $0.65$
& $-1.49$ \\
\hline
\end{tabular}%
\caption{
Marginalized $68\%$ C.L. constraints on $\Lambda\mathrm{LTB}$ density profile parameters ($m_{0,\mathrm{amp}}$, $r_0$) and model-comparison metrics relative to flat $\Lambda\mathrm{CDM}$ and $w_0w_a\mathrm{CDM}$. Results are derived by combining individual SNe~Ia compilations with DESI DR2 BAO and compressed Planck CMB data. We define $\Delta{\rm AIC}_{\Lambda{\rm CDM}}\equiv{\rm AIC}_{\Lambda{\rm LTB}}-{\rm AIC}_{\Lambda{\rm CDM}}$ and $\Delta\chi^2_{w_0w_a{\rm CDM}}\equiv\chi^2_{\Lambda{\rm LTB}}-\chi^2_{w_0w_a{\rm CDM}}$. The Bayesian evidence for model $M$ is $Z_{ M}\equiv P(d\mid M)$ (d is data vector), and the Bayes factor is $K_{\Lambda{\rm LTB},\Lambda{\rm CDM}}\equiv Z_{\Lambda{\rm LTB}}/Z_{\Lambda{\rm CDM}}$. Negative $\Delta\mathrm{AIC}$ and positive $\ln K$ favor $\Lambda\mathrm{LTB}$ over $\Lambda\mathrm{CDM}$ (strongest in DES SN5YR), while near-zero $\Delta\chi^2_{w_0w_a\mathrm{CDM}}$ indicates that $\Lambda\mathrm{LTB}$ and $w_0w_a\mathrm{CDM}$ yield comparable goodness-of-fit to distance measurements.
}
\label{tab:distance_constraints}
\end{table}

The above results rely on the assumption that the LTB geometry can be treated as a weak perturbation of the FRW metric. Appendix~\ref{app:distance_validation} tests the validity of this assumption using the most demanding posterior sample from the distance sector.

\begin{figure}[htbp]
\centering
\includegraphics[width=0.7\textwidth]{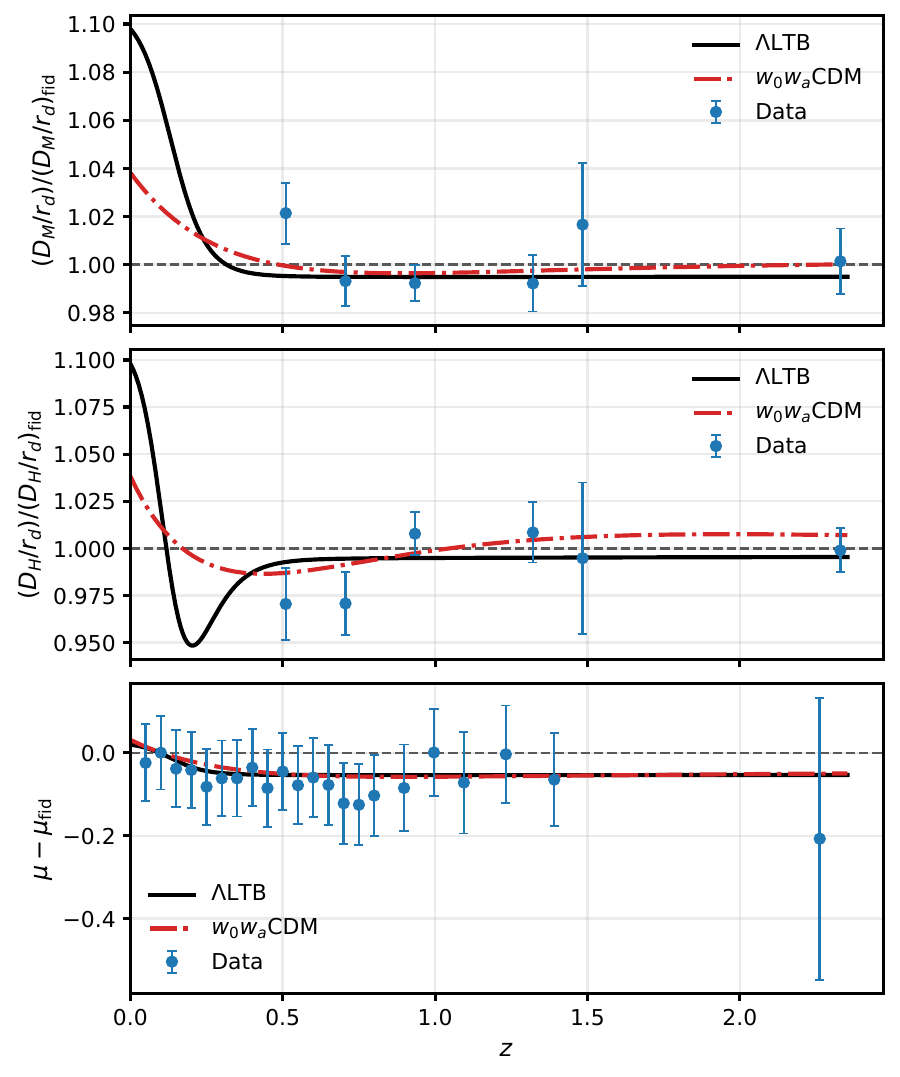}
\caption{
Distance--redshift relation for best-fit $\Lambda\mathrm{LTB}$ (black solid) and flat $w_0w_a\mathrm{CDM}$ (red dash-dotted) models relative to a flat $\Lambda\mathrm{CDM}$ reference, constrained by Union3 SNe~Ia + BAO + CMB. The top and middle panels show transverse $(D_{\mathrm{M}}/r_{\mathrm{d}})/(D_{\mathrm{M}}/r_{\mathrm{d}})_{\mathrm{fid}}$ and radial $(D_{\mathrm{H}}/r_{\mathrm{d}})/(D_{\mathrm{H}}/r_{\mathrm{d}})_{\mathrm{fid}}$ distance ratios with DESI BAO measurements (blue circles, $1\sigma$), respectively. The bottom panel shows distance modulus residuals $\mu - \mu_{\mathrm{fid}}$ (in~$\mathrm{mag}$) with Union3 SNe~Ia data. Compared to flat $\Lambda\mathrm{CDM}$, the $\Lambda\mathrm{LTB}$ model predicts pronounced low-redshift ($z \lesssim 0.5$) geometric deviations, whereas $w_0w_a\mathrm{CDM}$ exhibits milder evolution while yielding comparable SNe~Ia residuals.
}
\label{fig:distance_fit_comparison}
\end{figure}

Fig~\ref{fig:distance_fit_comparison} shows the distance-redshift relation ($D_M/r_d$, $D_H/r_d$, and $\mu$) for the best-fit $\Lambda\mathrm{LTB}$ and $w_0w_a\mathrm{CDM}$ models relative to $\Lambda$CDM, using the Union3 posterior sample. From the figure, $\Lambda\mathrm{LTB}$ fits all types of data better than $\Lambda$CDM, and is comparable to $w_0w_a\mathrm{CDM}$, in agreement with Table~\ref{tab:distance_constraints}. At low redshifts ($z\lesssim0.5$), $\Lambda\mathrm{LTB}$ exhibits a pronounced $z \to 0$ enhancement in both normalized $D_M/r_d$ and $D_H/r_d$ curves, along with a sharp dip in $D_H/r_d$ near $z \approx 0.2$.

\begin{figure}[htbp]
\centering
\includegraphics[width=0.7\textwidth]{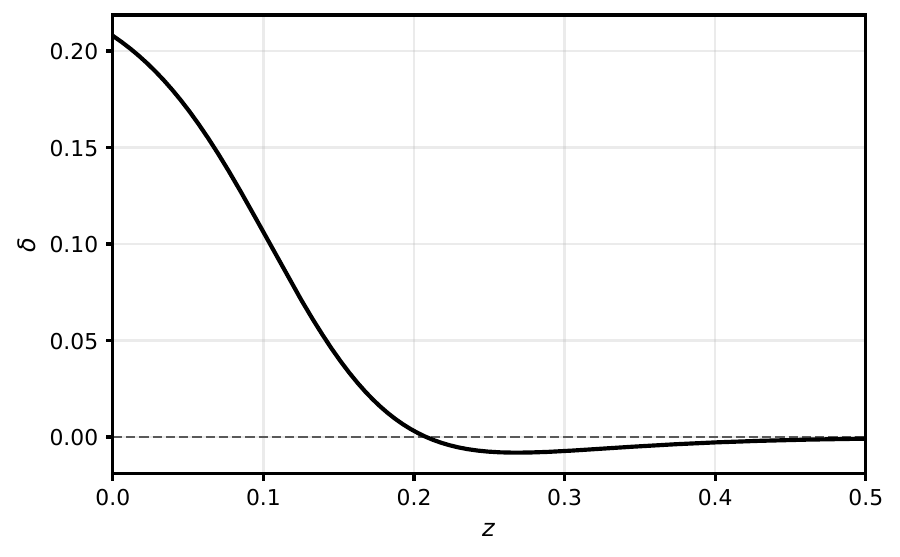}
\caption{
Local compensated density contrast $\delta(z)$ projected along the past light cone for the best-fit $\Lambda\mathrm{LTB}$ model constrained by Union3 $\mathrm{SNe~Ia} + \mathrm{BAO} + \mathrm{CMB}$. The black solid curve evolves from an overdense core with $\delta \approx 0.2$ at $z = 0$ to an underdense shell crossing zero at $z \approx 0.2$, and asymptotically approaches the homogeneous FRW background ($\delta = 0$, horizontal dashed line) by $z\sim0.5$. This spatial inhomogeneity induces expansion history mimicking dynamical dark energy equation of state that crosses $w = -1$ from $w>-1$ to $w<-1$ as redshift increases.
}
\label{fig:density_contrast}
\end{figure}

To clarify the radial structure underlying these low-redshift distance deviations, Fig.~\ref{fig:density_contrast} illustrates the local density contrast projected onto the past light cone. For the adopted compensated density profile, $\delta$ transitions from positive to negative, corresponding to a shift from an overdense to an underdense region. This spatial behavior resembles that of dynamical dark energy evolving from $w>-1$ to $w<-1$, crossing $w=-1$ as redshift increases, thus mimicking Quintom behavior.

\section{Constraining Radial Inhomogeneity with a kSZ Upper-bound Constraint}
\label{sec:ksz}

\subsection{LTB-induced kSZ signal}

The radial inhomogeneity not only alters light-cone distances but also drives a coherent radial velocity within the CMB rest frame. A larger profile amplitude increases this velocity, whereas a larger spatial scale maintains velocity coherence over larger radii. Consequently, the kSZ angular power spectrum is sensitive to both $m_{0,\mathrm{amp}}$ and $r_0$. To incorporate kSZ constraints, we transform the first-order $\Lambda\mathrm{LTB}$ solution into the conformal Newtonian gauge, in which standard kSZ calculations are naturally formulated \cite{Acoleyen2008}. We then compute the resulting kSZ angular power spectrum.

Because the perturbed LTB metric (Eq.~\ref{eq:conformal_ltb_distance}) and the conformal Newtonian gauge metric represent first-order perturbations around the same background FRW metric, they are related by a first-order coordinate transformation

\begin{equation}
\tilde\eta=\eta+T(\eta,r),
\qquad
\tilde r=r+L(\eta,r),
\qquad
T,L=\mathcal O(B,E_1),
\label{eq:coord_transform_ksz}
\end{equation}
so that the conformal Newtonian metric takes the form
\begin{equation}
ds^2=a^2(\tilde\eta)
\left[
-(1+2\Psi)d\tilde\eta^2
+
(1-2\Phi)(d\tilde r^2+\tilde r^2d\Omega^2)
\right].
\label{eq:CN_metric_ksz}
\end{equation}

Imposing the conformal Newtonian gauge conditions yields three equations. First, the vanishing shift condition, $\tilde{g}_{\tilde{\eta}\tilde{r}} = 0$, requires

\begin{equation}
\partial_\eta L=\partial_rT.
\label{eq:shift_condition_ksz}
\end{equation}
Spatial isotropy further requires
\begin{equation}
\Phi
=-\mathcal C+\mathcal HT+\partial_rL
=\mathcal HT+\frac{L}{r}-B.
\label{eq:Phi_two_forms_ksz}
\end{equation}
Equating the two expressions in Eq.~\eqref{eq:Phi_two_forms_ksz} yields a first-order radial equation for $L$,
\begin{equation}
\partial_rL-\frac{L}{r}
=\mathcal C-B
=r\,\partial_rB(\eta,r)-E_1(r).
\label{eq:L_radial_eq_ksz}
\end{equation}
Finally, the time-time component gives
\begin{equation}
\Psi=-(\mathcal HT+\partial_\eta T).
\label{eq:Psi_def_ksz}
\end{equation}
The residual zero mode is fixed by the center boundary conditions
\begin{equation}
T(\eta,0)=0,
\qquad
\lim_{r\to0}\frac{L(\eta,r)}{r}=B(\eta,0).
\label{eq:center_fix_BC_ksz}
\end{equation}
Under these conditions,
\begin{equation}
L(\eta,r)
=
r\left[
B(\eta,r)
-
\int_0^r
\frac{E_1(\bar r)}{\bar r}\,d\bar r
\right],
\qquad
T(\eta,r)
=
\int_0^r
\bar r\,\partial_\eta B(\eta,\bar r)\,d\bar r .
\label{eq:LT_solution_ksz}
\end{equation}
The corresponding Bardeen potentials are therefore
\begin{equation}
\Phi(\eta,r)
=
\mathcal H
\int_0^r
\bar r\,\partial_\eta B(\eta,\bar r)\,d\bar r
-
\int_0^r
\frac{E_1(\bar r)}{\bar r}\,d\bar r ,
\label{eq:Phi_ksz}
\end{equation}
\begin{equation}
\Psi(\eta,r)
=
-
\int_0^r
\bar r
\left[
\mathcal H\partial_\eta B(\eta,\bar r)
+
\partial_\eta^2B(\eta,\bar r)
\right]d\bar r .
\label{eq:Psi_ksz}
\end{equation}
Thus the two Bardeen potentials are fully determined. For a dust-plus-$\Lambda$ universe at first order, the traceless spatial Einstein equation yields $\Phi = \Psi$. Appendix~\ref{app:Phi_Psi_consistency} verifies that the LTB-to-Newtonian gauge transformation constructed here satisfies this identity. Although matter remains comoving in synchronous LTB coordinates, the corresponding worldlines acquire a non-zero peculiar velocity in conformal Newtonian gauge
\begin{equation}
v^r(\eta,r)
=
\frac{d\tilde r}{d\tilde\eta}
=
\partial_\eta L
=
r\,\partial_\eta B(\eta,r).
\label{eq:vr_expression_ksz}
\end{equation}
Appendix~\ref{app:velocity_accuracy} assesses the accuracy of this first-order velocity expression. With the potentials and radial peculiar velocity established, we compute the resulting kSZ angular power spectrum. The line-of-sight projected kSZ field is given by
\begin{equation}
\Theta(\hat{\mathbf n})
=
-
\int_0^{z_{\rm edge}}dz\,
g(z)\,
\beta_{\rm eff}(z)\,
\delta_e[D_M(z)\hat{\mathbf n},z].
\label{eq:Theta_projected_ksz}
\end{equation}
where the upper integration limit $z_{\mathrm{edge}}$ denotes the redshift beyond which radial density perturbations become negligible.
Here
\begin{equation}
g(z)=e^{-\tau(z)}\frac{d\tau}{dz}
\label{eq:visibility_ksz}
\end{equation}
is the visibility function, $\tau$ is the optical depth, and $\beta_{\mathrm{eff}}(z) \equiv v_{\mathrm{eff}}(z)/c$ denotes the effective velocity of free electrons at redshift $z$ relative to the CMB rest frame \cite{ZibinMoss2011}. The effective velocity field incorporates Doppler, Sachs--Wolfe (SW), and integrated Sachs--Wolfe (ISW) contributions. The Doppler term is evaluated from $v^r = r \partial_\eta B$ along the light cone, whereas the SW and ISW terms are integrated from $\Phi$ and $\Psi$ along the photon trajectory. The formulation of the three components employed in our pipeline is detailed in Appendix~\ref{app:ksz_kernel}.

For the post-reionization regime relevant to this work,
\begin{equation}
g(z)\simeq \frac{d\tau}{dz}.
\label{eq:g_approx_ksz}
\end{equation}
The optical depth from the observer to redshift $z$ is
\begin{equation}
\tau(z)
=
\int_0^z dz'\,
\sigma_T\,\bar n_e(z')\,
\left|\frac{c\,dt}{dz'}\right|,
\label{eq:tau_ksz}
\end{equation}
where $\sigma_T$ is the Thomson cross section. Along the central radial null geodesic,
\begin{equation}
\frac{dt}{dz}
=
-
\frac{1}{(1+z)H_\parallel(z)},
\qquad
H_\parallel(z)\equiv
\frac{\dot R'[t(z),r(z)]}{R'[t(z),r(z)]}.
\label{eq:dtdz_ksz}
\end{equation}
which leads to
\begin{equation}
\frac{d\tau}{dz}
=
\frac{\sigma_Tc\,\bar n_e(z)}
{(1+z)H_\parallel(z)}.
\label{eq:dtau_dz_ksz}
\end{equation}
The shell-averaged free-electron number density is modeled as
\begin{equation}
\bar n_e(z)
=
\frac{x_e(z)}
{m_p(1+4f_{\rm He})}\,
\bar\rho_b(z),
\label{eq:ne_ksz}
\end{equation}
where $x_e(z)$ is the effective ionization fraction, $m_{\mathrm{p}}$ is the proton mass, and $f_{\mathrm{He}} \equiv n_{\mathrm{He}}/n_{\mathrm{H}}$ is the helium-to-hydrogen number density ratio. The baryon density is assumed to trace the shell-averaged matter density at the background level,
\begin{equation}
\bar\rho_b(z)
=
\frac{\Omega_{b0}}{\Omega_{m0}}\,
\rho_m\!\left(a(\eta(z)),r(z)\right).
\label{eq:rhob_ksz}
\end{equation}
where $a(\eta(z))$ is evaluated via the first-order light-cone expansion in Eq.~\eqref{eq:a_lightcone_distance}.
For the first-order $\Lambda\mathrm{LTB}$ model considered here, the matter density evolves according to
\begin{equation}
\rho_m(a,r)
=
\frac{3H_0^2\Omega_{m0}}{8\pi G}\,
a^{-3}
\left[
1+
\left(
1-\frac{3}{2}H(a)F(a)
\right)\delta_0(r)
\right],
\label{eq:rhom_ksz}
\end{equation}
where $F(a)$ is defined in Eq.~\eqref{eq:B_solution_distance}, and $\delta_0(r)$ represents the present-day local density contrast. 

Equations~\eqref{eq:visibility_ksz}--\eqref{eq:rhom_ksz} specify $g(z)$, while the conformal Newtonian gauge construction determines $\beta_{\mathrm{eff}}$. Applying the flat-sky Limber approximation, the LTB-induced kSZ angular power spectrum is
\begin{equation}
C_\ell^{\rm kSZ}
\simeq
\int_0^{z_{\rm edge}}dz\,
\frac{H_\parallel(z)}
{cD_M^2(z)}
\left[
\frac{v_{\rm eff}(z)}{c}
\frac{d\tau}{dz}
\right]^2
P_e^{\rm FRW}
\left(
\frac{\ell+1/2}{D_M(z)},z
\right).
\label{eq:Clksz_ksz}
\end{equation}
In our baseline implementation, the electron power spectrum is modeled as
\begin{equation}
P_e^{\rm FRW}(k,z)
=
b_e^2(k,z)P_m^{\rm FRW}(k,z),
\qquad
b_e=1
\label{eq:Pe_ksz}
\end{equation}
Here, $P_m^{\mathrm{FRW}}(k,z)$ is the non-linear total matter power spectrum output by \texttt{CAMB} for the matching background FRW cosmology. High-precision emulators (e.g., the CSST cosmological emulator \cite{Chen2025CSSTEmulator}) can accelerate power spectrum evaluations with percent-level accuracy within their calibrated domains. However, at the high multipoles analyzed here ($\ell \sim 3000$), the low-redshift Limber wavenumber $k = (\ell + 1/2)/D_M(z)$ exceeds the emulator's calibrated domain, so we retain direct \texttt{CAMB} evaluations. In this baseline prescription, electron clustering is approximated directly by total matter clustering without introducing separate baryonic feedback or electron bias parameters. The kSZ signal amplitude is
\begin{equation}
D_\ell^{\rm kSZ}
=
T_{\rm CMB}^2
\frac{\ell(\ell+1)}{2\pi}
C_\ell^{\rm kSZ}.
\label{eq:Dell_ksz}
\end{equation}
Consequently, the same parameter vector in Eq.~\eqref{eq:theta_sampled_distance} simultaneously determines the distance observables and the LTB-induced kSZ power spectrum amplitude $D_{\ell=3000}^{\mathrm{kSZ}}$.

\subsection{Parameter constraints with a kSZ upper-bound constraint}

\begin{figure}[htbp]
\centering
\includegraphics[width=0.75\textwidth]{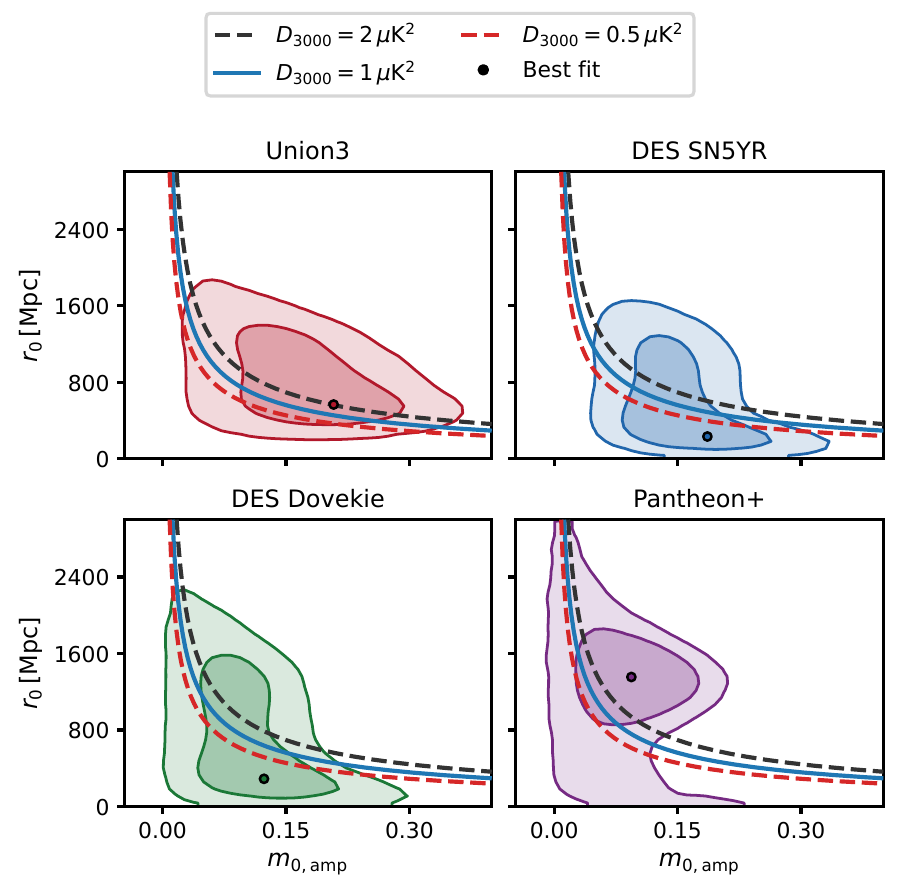}
\caption{
Marginalized 2D posterior distributions (68\% and 95\% C.L.) in the $(m_{0,\mathrm{amp}}, r_0)$ plane for the four SNe~Ia compilations, each combined with BAO and CMB data. Overlaid are curves of constant $D_{\ell=3000}^{\mathrm{kSZ}}$ ($0.5$, $1.0$, and $2.0\,\mu\mathrm{K}^2$). Circles denote best-fit points for each dataset. Regions above each curve represent kSZ-bright parameter space excluded by the corresponding $D_{\ell=3000}^{\mathrm{kSZ}}$ upper limits. Incorporating the kSZ constraint removes a substantial portion of the posterior probability, with the most stringent exclusion occurring for Pantheon+ and Union3.
}
\label{fig:sn_ksz_limit}
\end{figure}

To illustrate how the kSZ angular power spectrum constrains the density-profile amplitude and scale, Fig.~\ref{fig:sn_ksz_limit} overlays constant-$D_{\ell=3000}^{\rm kSZ}$ contours ($0.5$--$2\,\mu{\rm K}^2$) onto the distance-sector posterior distribution in the $(m_{0,\rm amp},r_0)$ plane.

For reference, the SPT-3G analysis reports the total kSZ angular power spectrum \cite{Chaubal2026} 
\begin{equation}
D_{3000}^{\rm kSZ,tot}
=
1.75\pm0.86\ \mu{\rm K}^2 .
\label{eq:spt_ksz_measurement}
\end{equation}
However, because the observed total incorporates standard homogeneous and patchy kSZ contributions, it cannot be attributed entirely to LTB-induced effects. Even a conservative upper limit of $D_{\ell=3000}^{\mathrm{kSZ}} = 2\,\mu\mathrm{K}^2$ excludes a substantial portion of the posterior probability. This restriction tightens systematically as the allowed residual power is reduced. Its impact is most pronounced for the Pantheon+ and Union3 posteriors, whereas it is comparatively weaker for DES SN5YR and DES Dovekie.

\begin{figure}[htbp]
\centering
\includegraphics[width=0.9\textwidth]{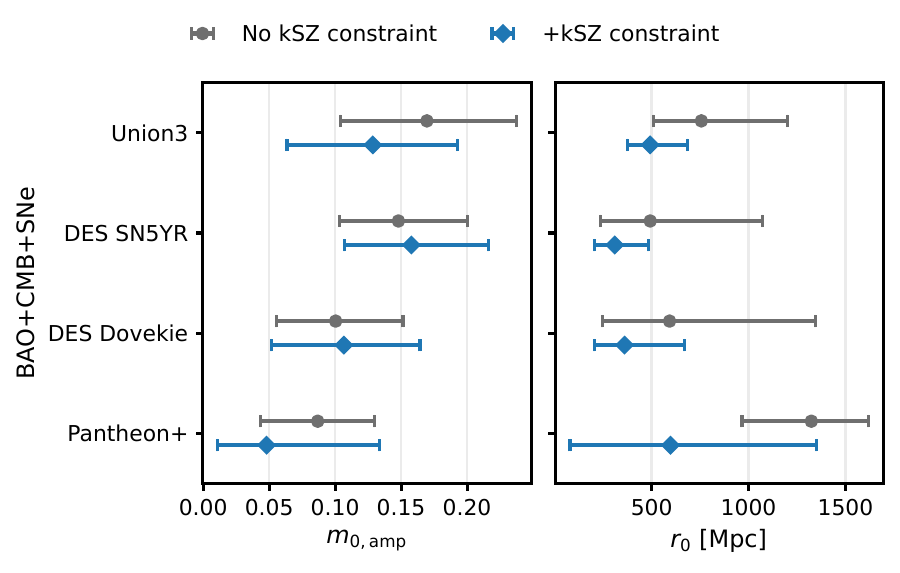}
\caption{
Comparison of marginalized 1D constraints (68\% C.L.) on $m_{0,\mathrm{amp}}$ (left) and $r_0$ (right) before (grey) and after (blue) imposing the kSZ upper limit $D_{\ell=3000}^{\mathrm{kSZ}} \leq 1\,\mu\mathrm{K}^2$. Applying the kSZ constraint shifts the amplitude $m_{0,\mathrm{amp}}$ toward smaller values primarily for Union3 and Pantheon+, whereas the DES constraints remain largely unaffected. In contrast, $r_0$ is systematically shifted toward smaller scales across all four cases, with its uncertainties substantially tightened for all analyses except Pantheon+.
}
\label{fig:sn_ksz_results}
\end{figure}

We implement the kSZ information as a one-sided external constraint. For each sampled parameter vector,
\begin{equation}
W_{\rm kSZ}(\Theta)
=
\mathcal H_{\rm step}
\left[
D_{\rm lim}
-
D_{\ell=3000}^{\rm kSZ}(\Theta)
\right],
\label{eq:ksz_constraint}
\end{equation}
where $W_{\rm kSZ}$ denotes the multiplicative weight assigned to the parameter vector $\Theta$, and $\mathcal H_{\rm step}$ is the Heaviside step function. The baseline analysis uses
\begin{equation}
D_{\rm lim}=1\,\mu{\rm K}^2 .
\label{eq:ksz_constraint_limit}
\end{equation}
Fig.~\ref{fig:sn_ksz_results} compares the marginalized $m_{0,\rm amp}$ and $r_0$ constraints before and after imposing this constraint. For $m_{0,\rm amp}$, the kSZ constraint has little effect on the DES SN5YR and DES Dovekie constraints, while it shifts the Union3 and Pantheon+ posteriors toward smaller amplitudes. For $r_0$, the constraint moves the allowed support across all four SNe~Ia combinations toward smaller characteristic radii, and substantially tightens the marginalized uncertainties for Union3, DES SN5YR, and DES Dovekie. For Pantheon+, the $D_{\ell=3000}^{\rm kSZ}=1\,\mu{\rm K}^2$ boundary removes a high-probability distance-sector region. The remaining allowed support is flatter in posterior weight, so the marginalized uncertainty does not simply shrink.

Appendix~\ref{app:D3000_fit} provides a fast fitting formula for estimating $D_{\ell=3000}^{\rm kSZ}$, which helps interpret how the kSZ angular power spectrum constrains the model parameters. Appendix~\ref{app:velocity_decomposition} presents the redshift dependence of the effective velocity field, including its Doppler, SW, and ISW components. All quantities are evaluated at the posterior medians derived from the four distance-sector joint analyses. Since this effective velocity is the main source of the LTB-induced kSZ angular power, it gives useful intuition for the relative kSZ amplitudes of the four cases.

Table~\ref{tab:ksz_d1_constraints} summarizes the results shown in Fig.~\ref{fig:sn_ksz_results} and reports the model-comparison metrics. Beyond the posterior shifts already visible in the figure, the table shows that the kSZ constraint has only a modest impact on $\Delta{\rm AIC}_{\Lambda{\rm CDM}}$, $\ln K_{\Lambda{\rm LTB},\Lambda{\rm CDM}}$, and $\Delta\chi^2_{w_0w_a{\rm CDM}}$.

\begin{table}[htbp]
\centering
\renewcommand{\arraystretch}{1.15}
\begin{tabular}{lccccc}
\hline\hline
\shortstack{SNe~Ia compilation\\(with BAO+CMB)}
& $m_{0,\rm amp}$
& $r_0$ (Mpc)
& $\Delta{\rm AIC}_{\Lambda{\rm CDM}}$
& \shortstack{Log Bayes factor\\$\ln K_{\Lambda{\rm LTB},\Lambda{\rm CDM}}$}
& $\Delta\chi^2_{w_0w_a{\rm CDM}}$ \\
\hline
Union3
& $0.129^{+0.064}_{-0.065}$
& $490^{+200}_{-120}$
& $-6.42$
& $0.82$
& $1.69$ \\
DES SN5YR
& $0.158^{+0.059}_{-0.051}$
& $310^{+170}_{-110}$
& $-11.59$
& $3.72$
& $0.07$ \\
DES Dovekie
& $0.107^{+0.058}_{-0.055}$
& $360^{+310}_{-160}$
& $-5.14$
& $0.84$
& $-0.27$ \\
Pantheon+
& $0.048^{+0.085}_{-0.037}$
& $600^{+750}_{-520}$
& $-2.61$
& $-0.58$
& $-0.20$ \\
\hline
\end{tabular}%
\caption{
Same as Table~\ref{tab:distance_constraints}, but with the external kSZ bound $D^\mathrm{kSZ}_{\ell=3000} \le 1\,\mu\mathrm{K}^2$. Imposing the kSZ constraint tightens parameter posterior and notably shrinks the scale radius $r_0$, yet it has only a modest impact on overall model metrics. Specifically, DES SN5YR maintains strong support for $\Lambda\mathrm{LTB}$ over $\Lambda\mathrm{CDM}$, while $\Lambda\mathrm{LTB}$ yields a goodness-of-fit comparable to $w_0w_a\mathrm{CDM}$.
}
\label{tab:ksz_d1_constraints}
\end{table}

The three model-comparison metrics remain nearly unchanged for DES SN5YR and DES Dovekie. For Union3, the log Bayes factor shows a weaker preference for $\Lambda\mathrm{LTB}$, while the remaining metrics remain largely unchanged. In contrast, Pantheon+ exhibits a consistently reduced preference for $\Lambda\mathrm{LTB}$ across all three metrics. Nevertheless, these shifts do not alter our primary conclusion. The data broadly favor $\Lambda\mathrm{LTB}$ over $\Lambda\mathrm{CDM}$; the $\Lambda$LTB and $w_0w_a$CDM models describe the data almost equally well. The $\Lambda$LTB fit is slightly poorer, but the two models remain statistically consistent.

As a robustness check, Appendix~\ref{app:width_robustness} repeats the inference pipeline adopting a narrower transition-width ratio of $\Delta r/r_0 = 0.2$ (in place of the baseline value $0.4$). The resulting parameter constraints and model-comparison metrics change little.

\subsection{Implications for the Copernican principle}

The kSZ constraint primarily excludes the high-velocity parameter space induced by large $m_{0,\rm amp}$ or $r_0$. Consequently, the surviving radial profiles align more closely with expectations for Gaussian fluctuations in standard $\Lambda\mathrm{CDM}$. To quantify this behavior, adapting the approach of Refs.~\cite{Valkenburg2014,Camarena2021}, we compare the inferred enclosed density perturbation with the linear $\Lambda\mathrm{CDM}$ density variance on the corresponding scale. For a spherical top-hat window of radius $R$, the variance at $z=0$ is

\begin{equation}
\sigma_R^2
=
\frac{1}{2\pi^2}
\int_0^\infty dk\,k^2P_m(k,z=0)
\left[
\frac{3j_1(kR)}{kR}
\right]^2,
\label{eq:sigma_R_copernican}
\end{equation}
where $P_m(k,z=0)$ is the matter power spectrum of the matched $\Lambda\mathrm{CDM}$ background and $j_1$ is the spherical Bessel function of the first kind. In the linear, large-scale limit, the mean density perturbation within the sphere is approximately Gaussian,
\begin{equation}
\delta_R\sim\mathcal N(0,\sigma_R^2).
\label{eq:delta_R_gaussian}
\end{equation}
where $\delta_R$ denotes the mean density perturbation enclosed within a radius $R$, corresponding to $m(R)$ defined in Eq.~\eqref{eq:enclosed_delta_distance}. We therefore define
\begin{equation}
\nu(R)\equiv\frac{m(R)}{\sigma_R},
\label{eq:nu_copernican}
\end{equation}
which quantifies the deviation of the inferred region from the Copernican expectation.

\begin{figure}[htbp]
\centering
\includegraphics[width=0.8\textwidth]{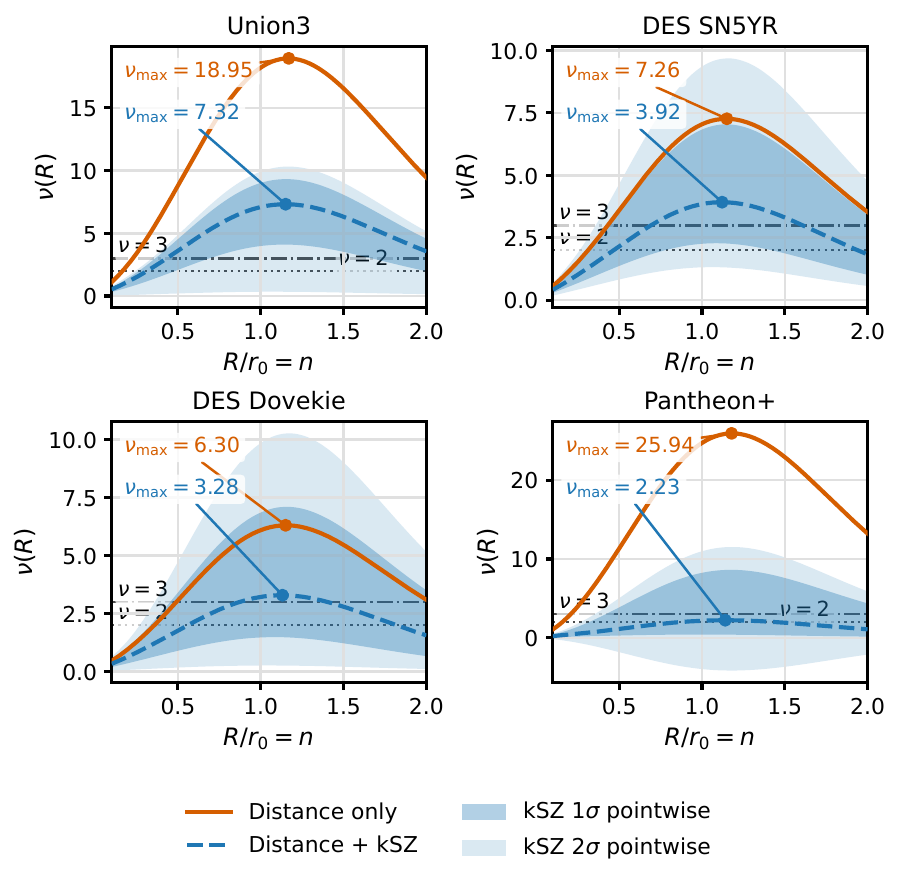}
\caption{
Statistical significance $\nu(R) = m(R)/\sigma_R$ of the enclosed density deviation relative to the Copernican expectation, plotted against normalized radius $n = R/r_0$. Medians before (solid orange) and after (dashed blue) imposing the kSZ constraint are shown, along with $1\sigma$ and $2\sigma$ pointwise credible bands for the kSZ-constrained posteriors. Annotations show the median peak values $\nu_{\mathrm{max}}$, while horizontal lines mark the $\nu=2$ and $\nu=3$ significance thresholds. Although the kSZ constraint substantially suppresses $\nu(R)$, the $1\sigma$ lower bounds for Union3 and the DES datasets remain above $\nu = 2\text{--}3$ at their peaks. In contrast, Pantheon+ yields a median peak of $\nu_{\mathrm{max}} = 2.23$ (with its $1\sigma$ band encompassing $\nu \le 2$), demonstrating consistency with the Copernican expectation. At the $2\sigma$ level, all four analyses remain fully compatible with the Copernican principle.
}
\label{fig:copernican_prior}
\end{figure}

Fixing the primordial power spectrum parameters to $A_s=2.1\times10^{-9}$ and $n_s=0.965$, we evaluate $\nu(R)$ over the dimensionless radius $n\equiv R/r_0\in[0.1,2]$. Fig.~\ref{fig:copernican_prior} displays the posterior median for the distance-only samples alongside the median and $1\sigma$/$2\sigma$ credible bands obtained after imposing the kSZ constraint.

Imposing the kSZ constraint substantially suppresses $\nu(R)$ across all four distance analyses. The remaining departure from the Copernican expectation differs among the SNe~Ia compilations. For Union3, the lower boundary of the $1\sigma$ credible band at the peak remains above $\nu=3$, indicating strong tension with the standard $\Lambda$CDM prediction. For DES SN5YR, the corresponding lower boundary remains above $\nu=2$, so the tension is weaker but still significant. For DES Dovekie, the peak region of the $1\sigma$ band lies mostly above $\nu=2$, indicating a tendency to depart from the $\Lambda$CDM expectation. The Pantheon+ posterior remains broadly consistent with the Copernican principle. In all four analyses, the lower boundary of $2\sigma$ credible band at the peak includes $\nu=2$, so compatibility with the Copernican principle cannot be excluded at high confidence.
\section{Discussion and Conclusion}
\label{sec:discussion}

In this work, we develop a first-order perturbative framework for the $\Lambda\mathrm{LTB}$ model. We then apply this framework to test whether large-scale radial inhomogeneity can serve as an alternative to dynamical dark energy. We find that the $\Lambda\mathrm{LTB}$ model fits current distance data comparably well to $w_0w_a\mathrm{CDM}$, while kSZ constraints substantially restrict, but do not rule out, the viable parameter space.

The preferred profile is qualitatively consistent with the DESI DR2 DDE result \cite{DESIDR2ResultsII}, where $w_0 > -1$ modifies the low-redshift distance--redshift relation in a manner analogous to a local overdensity. Because our compensated profile transitions from an overdensity to an underdensity with increasing radius, it can further mimic an apparent Quintom behavior \cite{feng2005dark,ren2026short,MarraPaakkonen2010,RomanoChen2011,RomanoSasakiStarobinsky2012}. The kSZ constraint excludes profiles with both large amplitudes and large characteristic scales, substantially reducing the allowed departure from the Copernican principle. However, because the surviving low-kSZ profiles retain nearly the same peak likelihood, the minimum $\chi^2$ and AIC remain largely unchanged. Instead, the kSZ bound primarily reduces the high-likelihood posterior volume, weakening the Bayes factor for some SNe~Ia combinations without altering the qualitative model preference.

Several recent $\Lambda\mathrm{LTB}$ analyses have examined whether a radial inhomogeneity can alleviate the Hubble tension (for a review of the Hubble tension, see~\cite{cai2026hubble}). Most of these studies \cite{HoscheitBarger2018,Ding2019,Lukovic2020,Cai2021,Camarena2022} favor an underdense profile, in contrast to our findings. This discrepancy is largely driven by their inclusion of a SH0ES-like local $H_0$ prior, whereas our analysis does not impose this prior. Furthermore, large-scale underdensities are generally insufficient to resolve the Hubble tension \cite{Castello2022}. Without assuming an $H_0$ prior, a recent analysis combining DESI DR2 BAO, Planck CMB, and Pantheon+/DES Y5 SNe~Ia data also favors an overdense profile \cite{Camarena2025}. Both that study and the present work test whether large-scale radial inhomogeneities can account for distance--redshift signatures typically attributed to dynamical dark energy. In contrast, Futamase et al. conducted a distinct yet complementary analysis \cite{Futamase2026}. Specifically, they interpreted the local $H_0$ tension and the DESI DR2 BAO deviations from the standard cosmology as possible signatures of cosmic inhomogeneities, using the combined dataset to reconstruct a redshift-dependent matter density profile. In addition to matter density inhomogeneities, radial variations in dark-energy density arising from late-time first-order phase transitions have also been explored \cite{pen2014observational,friedman2026dark}. Recently, Friedman-Shaw et al. \cite{friedman2026dark} showed that the resulting dark-energy bubbles can produce distance--redshift signatures similar to DESI BAO data. However, the parameter region capable of producing such signatures is ruled out by CMB constraints, particularly those from kSZ measurements.

Beyond these observational considerations, our perturbative analytic framework offers a computationally efficient alternative for solving the LTB equations, avoiding the use of implicit elliptic integrals. While symbolic regression has recently been used to construct closed-form LTB approximations \cite{Carvalho2026}, our derivation keeps the underlying physical connections among the metric, gravitational potentials, and velocity field fully explicit. Formulating the solution in conformal Newtonian gauge also ensures compatibility with the standard cosmological perturbation framework. This enables straightforward extensions to observables sensitive to velocity and gravitational potentials, such as redshift-space distortions, Doppler magnification, and gravitational redshift. Furthermore, this gauge formulation substantially streamlines the kSZ calculation. For off-center sources in LTB spacetimes, traditional kSZ calculations integrate inward and outward null geodesics from each scattering location to the last scattering surface. The resulting boundary redshifts ($z^*_+$ and $z^*_-$) are then used to determine the effective velocity \cite{GarciaBellido2008,Yoo2010,Bull2012}. In contrast, our method evaluates the effective velocity directly from the potentials and velocity field in the conformal Newtonian gauge. This bypasses the need for repeated numerical integrations of null geodesics and yields an explicit decomposition into Doppler, Sachs--Wolfe, and integrated Sachs--Wolfe contributions.

The present analysis has several limitations. First, exact spherical symmetry is an idealization of large-scale inhomogeneities, and the adopted density profile represents a restricted functional family. Second, while the first-order $\Lambda\mathrm{LTB}$ framework is sufficiently accurate within the parameter space allowed by the current data, it breaks down when the metric departs significantly from a FRW spacetime. Third, we approximate the electron power spectrum in the LTB spacetime using the matter power spectrum of a matched FRW background. This simplification neglects spacetime-inhomogeneity effects on clustering and electron bias, as well as small-scale nonlinear physics. Future work should refine the kSZ modeling, incorporate more general or physically motivated profiles, and relax the spherical symmetry assumption to accommodate three-dimensional large-scale inhomogeneities.

Overall, our results demonstrate that large-scale inhomogeneity remains a viable alternative to dynamical dark energy for interpreting distance–redshift signatures in current data. Although the kSZ constraint substantially restricts allowed departures from homogeneity, it does not change the above conclusion. Meanwhile, the inferred deviation from the Copernican principle remains dataset-dependent. To further distinguish these two interpretations, future analyses could exploit the complementary information on cosmic expansion and structure growth provided by large-scale structure probes (for a review, see~\cite{zhang2026large}). The first-order $\Lambda\mathrm{LTB}$ framework developed here provides a unified, computationally efficient foundation for future modeling, facilitating tests of large-scale inhomogeneity against complementary cosmological probes.


\clearpage
\appendix

\section{Consistency of the Bardeen potentials}
\label{app:Phi_Psi_consistency}

The Bardeen potentials derived in Eqs.~\eqref{eq:Phi_ksz} and \eqref{eq:Psi_ksz} can be written as
\begin{equation}
\Phi(a,r)
=
\frac{\mathcal J(r)}{2}
\left[
a^3H^2\partial_aS+\alpha
\right],
\label{eq:app_Phi_initial}
\end{equation}
and
\begin{equation}
\Psi(a,r)
=
-\frac{\mathcal J(r)}{2}
\left[
a^3H^2\partial_aS
+a^2H\partial_a\left(a^2H\partial_aS\right)
\right],
\label{eq:app_Psi_initial}
\end{equation}
where
\begin{equation}
\alpha\equiv\frac{I_m(1)}{I_E(1)},
\qquad
S(a)\equiv H(a)F(a),
\qquad
\mathcal J(r)\equiv\int_0^r d\bar r\,\bar r\,m(\bar r).
\label{eq:app_alpha_S_J}
\end{equation}
Subtracting Eq.~\eqref{eq:app_Phi_initial} from Eq.~\eqref{eq:app_Psi_initial} and combining the time derivatives yields

\begin{equation}
\Psi-\Phi
=
-\frac{\mathcal J(r)}{2}
\left[
H\partial_a\left(a^4H\partial_aS\right)+\alpha
\right].
\label{eq:app_Psi_minus_Phi}
\end{equation}
Therefore, the equality $\Phi = \Psi$ is equivalent to
\begin{equation}
\partial_a\left(a^4H\partial_aS\right)
=
-\frac{\alpha}{H}.
\label{eq:app_key_identity}
\end{equation}

From the definitions of $F(a)$, $I_m(a)$, and $I_E(a)$ in Eqs.~\eqref{eq:Im_IE_distance} and \eqref{eq:B_solution_distance},
\begin{equation}
S(a)=H(a)I_m(a)-\alpha H(a)I_E(a),
\label{eq:app_S_split}
\end{equation}
with
\begin{equation}
\partial_a I_m=\frac{\Omega_m}{aH},
\qquad
\partial_a I_E=\frac{1}{a^3H^3}.
\label{eq:app_Im_IE_derivatives}
\end{equation}
For the spatially flat dust plus $\Lambda$ background in Eq.~\eqref{eq:bg_H_distance},
\begin{equation}
\partial_a\left(a^4H\partial_aH\right)=0,
\qquad
\partial_a\Omega_m
=
\Omega_m\left(
-\frac{3}{a}-2\frac{\partial_aH}{H}
\right).
\label{eq:app_background_identities}
\end{equation}
Applying these relations to the two terms in Eq.~\eqref{eq:app_S_split} yields
\begin{equation}
\partial_a\left[
a^4H\partial_a(HI_m)
\right]=0,
\qquad
\partial_a\left[
a^4H\partial_a(HI_E)
\right]=\frac{1}{H}.
\label{eq:app_two_identities}
\end{equation}
Substitution into Eq.~\eqref{eq:app_S_split} immediately gives Eq.~\eqref{eq:app_key_identity}. Hence $\Psi-\Phi=0$, as required for pressureless matter plus $\Lambda$ when anisotropic stress is neglected.

\begin{figure}[htbp]
\centering
\includegraphics[width=0.8\textwidth]{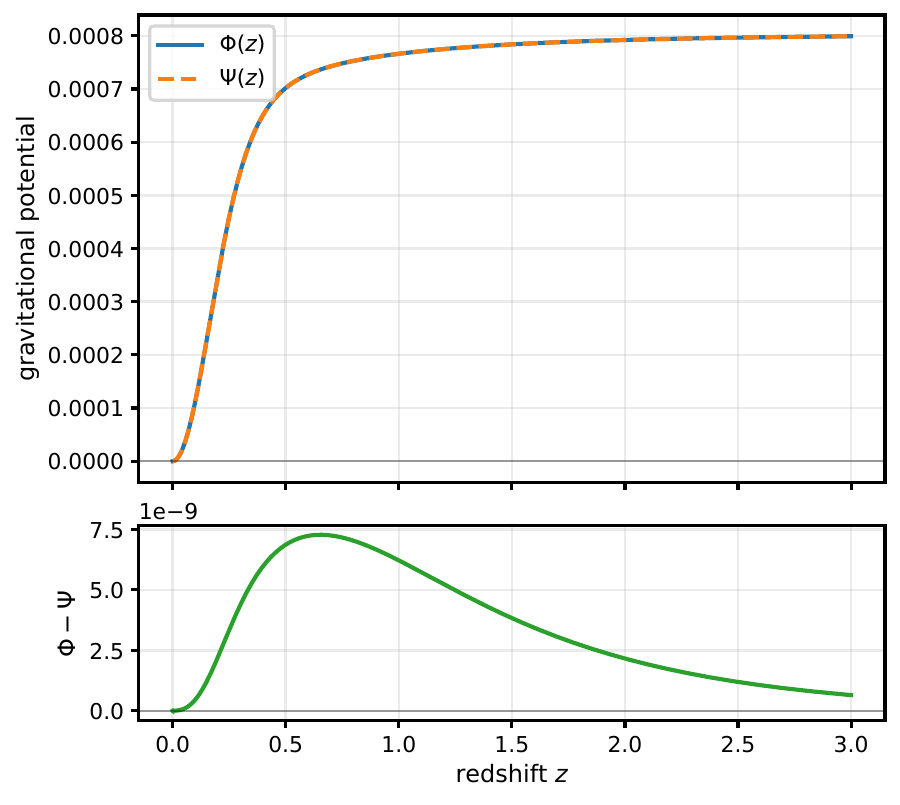}
\caption{
Redshift evolution of the Bardeen potentials $\Phi(z)$ and $\Psi(z)$, evaluated at the Union3 distance-only posterior medians. The upper panel shows $\Phi(z)$ (solid blue) and $\Psi(z)$ (dashed orange), which are visually indistinguishable across the entire redshift range $z \in [0, 3]$. The lower panel displays their difference, $\Phi - \Psi$, which peaks at $\sim 8 \times 10^{-9}$. This residual is approximately five orders of magnitude smaller than the individual potential amplitudes ($\sim 8 \times 10^{-4}$), confirming the analytical identity $\Phi = \Psi$ to numerical precision.
}
\label{fig:app_phi_psi_numerical}
\end{figure}

We further verify this equality numerically. Fig.~\ref{fig:app_phi_psi_numerical} illustrates the redshift evolution of $\Phi$ and $\Psi$, evaluated at the posterior-median parameters from the Union3 distance-only analysis.

The difference between the two potentials is approximately five orders of magnitude smaller than their individual amplitudes. This numerical agreement confirms the analytic identity $\Phi=\Psi$.

\section{kSZ effective-velocity}
\label{app:ksz_kernel}

Following Refs.~\cite{ZhangJohnson2015,Terrana2017}, the effective CMB dipole observed by an electron at $\chi_e \equiv \chi(z_e)$ can be written as
\begin{equation}
\beta_{\rm eff}(\hat{\mathbf n}_e,z_e)
=
i\int\frac{d^3k}{(2\pi)^3}
\Psi_{\rm m}(\mathbf k)
\mathcal K^v(k,z_e)
P_1(\hat{\mathbf k}\cdot\hat{\mathbf n}_e)
e^{i\chi_e\mathbf k\cdot\hat{\mathbf n}_e},
\label{eq:app_beta_eff_general}
\end{equation}
where $P_1(\mu)=\mu$. Here, the reference potential $\Psi_{\rm m}$ represents the asymptotic matter-dominated era potential in a dust plus $\Lambda$ background
\begin{equation}
\Psi_{\rm m}(\mathbf x)
\equiv
\lim_{a'\to0}\Psi(a',\mathbf x),
\qquad
\Psi_{\rm m}(\mathbf k)
=
\int d^3x\,
e^{-i\mathbf k\cdot\mathbf x}
\Psi_{\rm m}(\mathbf x).
\label{eq:app_Psi_m_FT}
\end{equation}
The total velocity kernel is decomposed as
\begin{equation}
\mathcal K^v(k,z_e)
=
\mathcal K_{\rm D}(k,z_e)
+\mathcal K_{\rm SW}(k,z_e)
+\mathcal K_{\rm ISW}(k,z_e),
\label{eq:app_Kv_split}
\end{equation}
with
\begin{align}
\mathcal K_{\rm D}(k,z_e)
&=
kD_v(a_{\rm dec})j_0(k\Delta\chi_{\rm dec})
-2kD_v(a_{\rm dec})j_2(k\Delta\chi_{\rm dec})
-kD_v(a_e),
\label{eq:app_K_D}
\\
\mathcal K_{\rm SW}(k,z_e)
&=
3\left[
2D_\Psi(a_{\rm dec})-\frac{5}{3}
\right]
j_1(k\Delta\chi_{\rm dec}),
\label{eq:app_K_SW}
\\
\mathcal K_{\rm ISW}(k,z_e)
&=
6\int_{a_{\rm dec}}^{a_e}da\,
\frac{dD_\Psi}{da}
j_1[k\Delta\chi(a)].
\label{eq:app_K_ISW}
\end{align}
The coefficient $5/3$ in Eq.~\eqref{eq:app_K_SW} arises from the reference-potential convention adopted here.\footnote{Standard derivations normalize transfer functions to the primordial potential during the radiation-dominated era. Because the present LTB construction contains no radiation, we normalize instead to the asymptotic matter-era potential $\Psi_{\rm m}$. This re-scaling modifies the prefactor in the standard SW kernel from $3/2$ to $(3/2)/(9/10) = 5/3$.} The relevant scale factors and distance variables are
\begin{equation}
a_e=(1+z_e)^{-1},
\qquad
\Delta\chi(a)=\chi(a)-\chi_e,
\qquad
\Delta\chi_{\rm dec}=\chi(a_{\rm dec})-\chi_e,
\label{eq:app_delta_chi}
\end{equation}
where $a_{\rm dec}$ is the scale factor at photon decoupling, and $\chi(a)$ is the background comoving distance from scale factor $a$ to the present epoch
\begin{equation}
\chi(a)=\int_a^1\frac{d\tilde a}{\tilde a^2H(\tilde a)}.
\label{eq:app_chi}
\end{equation}
Replacing these background relations with their first-order LTB counterparts affects the effective velocity only at second order. Here, $j_\ell$ denote spherical Bessel functions, while $D_\Psi$ and $D_v$ are the growth functions for the potential and velocity, derived respectively from the potential $\Psi$ and the radial peculiar velocity $v^r$.

Using the first-order solution, the two potentials take the separable form
\begin{equation}
\Psi(a,r)
=
\frac{P(a)}{2}\mathcal J(r)
=
\Phi(a,r),
\label{eq:app_Psi_PJ}
\end{equation}
with
\begin{equation}
\mathcal J(r)
\equiv
\int_0^r d\bar r\,\bar r\,m(\bar r),
\label{eq:app_J_kernel}
\end{equation}
where the additive large-radius mode is removed during numerical transformations.\footnote{Provided $m(r)$ decays rapidly enough at large radii, $\mathcal{J}(r)$ asymptotically approaches a finite constant $\mathcal{J}_\infty$. Setting the central potential to zero causes $\Psi_{\rm m}(r)$ to approach an additive constant. Although this constant does not influence peculiar velocities or potential gradients for $k>0$, it introduces a pure $k=0$ mode that can contaminate numerical Fourier transforms. We therefore eliminate this zero mode in our numerical implementation.}

The time-dependent factor is
\begin{equation}
P(a)
\equiv
a^3H^2(a)\frac{d[H(a)F(a)]}{da}
+\frac{I_m(1)}{I_E(1)}
=
\frac{\Omega_{m0}H_0^2}{a}
\left[
1-\frac{3}{2}H(a)F(a)
\right].
\label{eq:app_P_definition}
\end{equation}
Thus the potential growth factor can be written as
\begin{equation}
D_\Psi(a)
=
\frac{\Psi(a,r)}{\lim_{a'\to0}\Psi(a',r)}
=
\frac{P(a)}{\lim_{a'\to0}P(a')}.
\label{eq:app_D_Psi}
\end{equation}
The radial peculiar velocity in Eq.~\eqref{eq:vr_expression_ksz} can be rewritten as
\begin{equation}
v^r
=
r\partial_\eta B
=
\frac{a^2H(a)}{2}
\frac{d[H(a)F(a)]}{da}
r\,m(r)
=
-D_v(a)\partial_r\Psi_{\rm m}(r),
\label{eq:app_vr_growth}
\end{equation}
which gives
\begin{equation}
D_v(a)
=
-\frac{a^2H(a)}{\lim_{a'\to0}P(a')}
\frac{d[H(a)F(a)]}{da}.
\label{eq:app_D_v_direct}
\end{equation}
Defining the density growth function and growth rate as
\begin{equation}
D_\delta(a)
\equiv
1-\frac{3}{2}H(a)F(a),
\qquad
f(a)
\equiv
\frac{d\ln D_\delta}{d\ln a},
\label{eq:app_D_delta_f}
\end{equation}
the velocity growth factor can also be written as
\begin{equation}
D_v(a)
=
\frac{2f(a)}{3\Omega_{m0}H_0^2}
a^2H(a)D_\Psi(a).
\label{eq:app_D_v_growth}
\end{equation}
thus all quantities entering Eq.~\eqref{eq:app_beta_eff_general} are determined.

For a general perturbation, $\beta_{\rm eff}$ depends on both direction and redshift. However, because the profiles considered in this work are spherically symmetric about the central observer, the potential simplifies to $\Psi_{\rm m}(\mathbf{k}) = \Psi_{\rm m}(k)$. Substituting
\begin{equation}
\int d\Omega_k\,
P_1(\hat{\mathbf k}\cdot\hat{\mathbf n})
e^{ik\chi\hat{\mathbf k}\cdot\hat{\mathbf n}}
=
4\pi i\,j_1(k\chi),
\label{eq:app_angular_integral}
\end{equation}
each contribution reduces to the one-dimensional integral
\begin{equation}
\beta_X(z)
=
-\frac{1}{2\pi^2}
\int_0^\infty dk\,k^2
\Psi_{\rm m}(k)
\mathcal K_X(k,z)
j_1[k\chi(z)],
\qquad
X\in\{\mathrm D,\mathrm{SW},\mathrm{ISW}\}.
\label{eq:app_beta_X}
\end{equation}
The total effective velocity source is therefore given by
\begin{equation}
\beta_{\rm eff}(z)
=
\beta_{\rm D}(z)
+\beta_{\rm SW}(z)
+\beta_{\rm ISW}(z).
\label{eq:app_beta_sum}
\end{equation}

\section{Accuracy of the first-order perturbation}
\label{app:distance_validation}

\begin{figure}[htbp]
\centering
\includegraphics[width=0.8\textwidth]{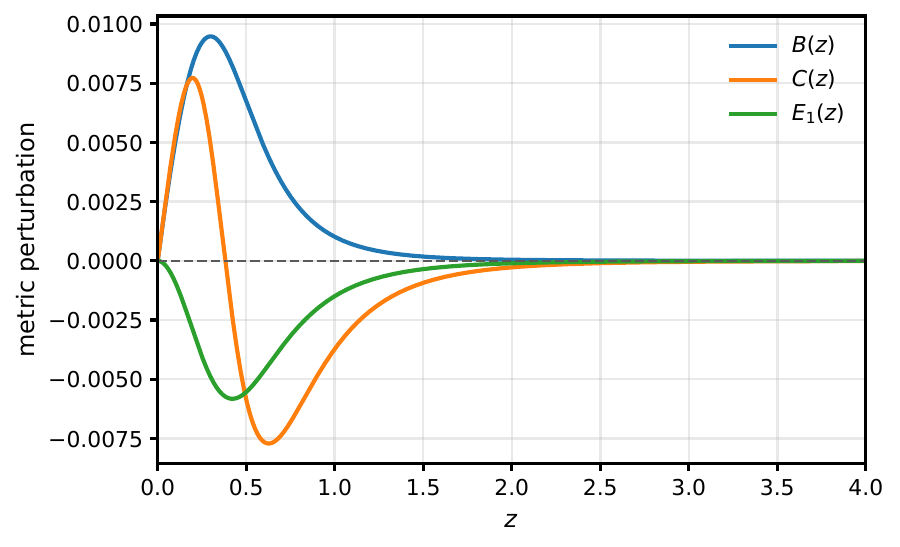}
\caption{
Stress test of the first-order linear approximation, showing the dimensionless metric perturbations evaluated along the central past light cone, up to $z = 4$. The test uses the posterior sample exhibiting the largest peak areal-radius perturbation ($\max_z |B(z)|$), across all distance-only analyses. Shown are the perturbations in areal radius $B(z)$ (blue), radial null propagation $C(z) = B + r\partial_r B - E_1$ (orange), and curvature $E_1(z)$ (green). All three perturbations remain strictly bounded by $\lesssim 0.01$ ($\ll 1$), confirming that the metric perturbations stay well within the linear regime, even for the most extreme posterior sample.
}
\label{fig:app_metric_stress}
\end{figure}

In the baseline analysis, we treat the LTB metric as a first-order perturbation around a spatially flat FRW background. To stress-test this approximation, we select the posterior sample that maximizes the peak areal-radius perturbation, $\max_z|B(z)|$, across the distance-only analyses of the four SNe~Ia compilations. We use this sample solely for validation and plot its corresponding metric perturbations in Fig.~\ref{fig:app_metric_stress}.

For this sample,
\begin{equation}
\max_z|B(z)|=0.0095
\qquad
\max_z|\mathcal C(z)|=0.0077,
\qquad
\max_z|E_1(z)|=0.0058.
\label{eq:app_metric_maxima}
\end{equation}
the metric perturbations remain small, confirming that the linear approximation holds even for the posterior sample with the largest peak areal-radius perturbation $\max_z|B(z)|$.

We further test the approximation on the distance--redshift relation, comparing the first-order areal radius to a full numerical LTB solution for the same profile.

\begin{figure}[htbp]
\centering
\includegraphics[width=0.75\textwidth]{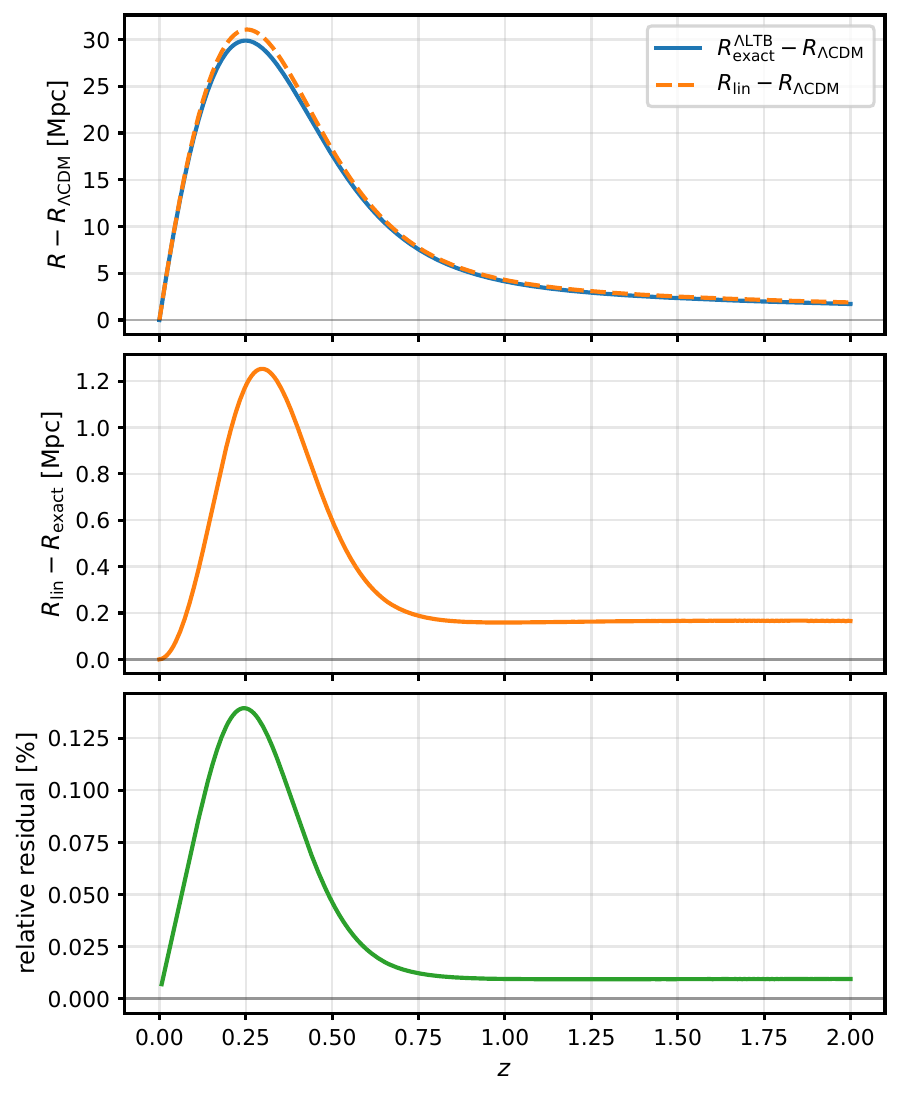}
\caption{
Precision test of the first-order areal radius $R_{\mathrm{lin}} = ar(1+B)$ against the exact numerical LTB solution $R_{\mathrm{exact}}^{\Lambda\mathrm{LTB}}$, for the stress-test profile (Fig. 6). Top panel: Perturbative deviations from the flat $\Lambda\text{CDM}$ background ($R - R_{\Lambda\text{CDM}}$), comparing the full numerical solution (blue solid) and the first-order approximation (orange dashed). Middle panel: Absolute error $R_{\mathrm{lin}} - R_{\mathrm{exact}}$, which peaks at $\sim 1.25\,\mathrm{Mpc}$ near $z \sim 0.3$. Bottom panel: Relative fractional residual $(R_{\mathrm{lin}} - R_{\mathrm{exact}})/R_{\mathrm{exact}}$, reaching a maximum of only $\approx 0.14\%$. This sub-percent accuracy confirms that the linear approximation introduces negligible theoretical error, remaining far below observational uncertainties.
}
\label{fig:app_R_validation}
\end{figure}

Figures~\ref{fig:app_metric_stress} and \ref{fig:app_R_validation} present these complementary checks. Because the metric perturbations remain small throughout the relevant posterior region, the resulting error in the distance--redshift relation is negligible. Consequently, the first-order treatment offers sufficient accuracy for the present analysis.

\section{Accuracy of the first-order velocity expression}
\label{app:velocity_accuracy}

Eq.~\eqref{eq:vr_expression_ksz} provides a first-order expression for the radial peculiar velocity in the conformal Newtonian gauge. Because both the LTB metric and the coordinate transformation are truncated at linear order, it's hard to retain a complete set of second-order terms. To assess accuracy, we restore the terms omitted when reducing $d\tilde r/d\tilde\eta$ to $\partial_\eta L$ in Eq.~\eqref{eq:vr_expression_ksz}. These include the Jacobian factor $d\tilde\eta/d\eta$ and the pullback between $(\tilde\eta,\tilde r)$ and $(\eta,r)$.
\begin{figure}[htbp]
\centering
\includegraphics[width=0.8\textwidth]{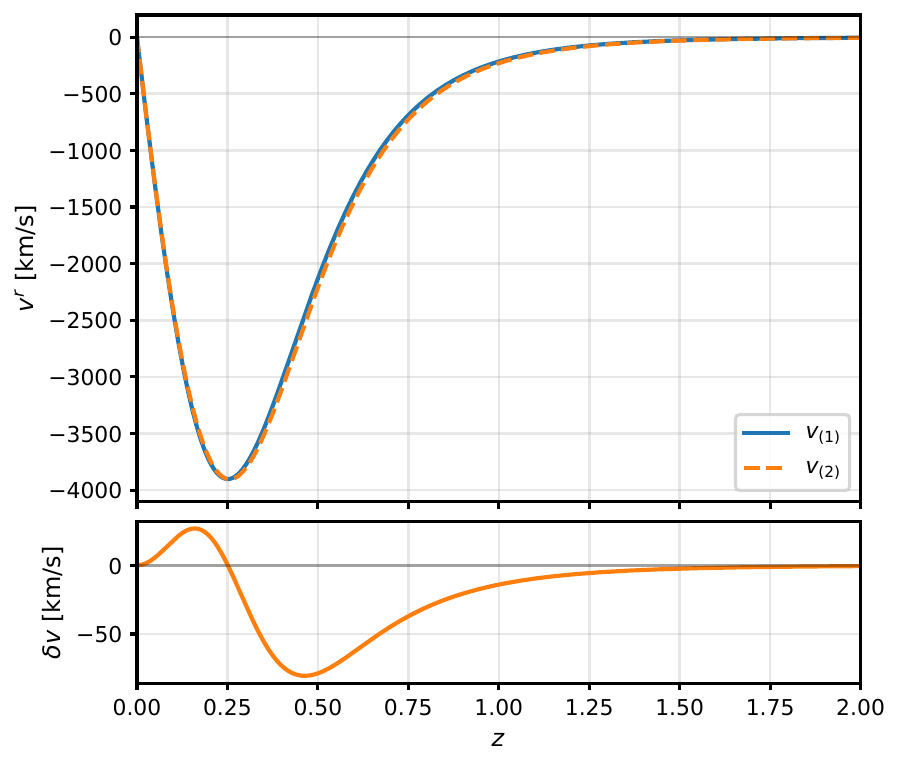}
\caption{
Accuracy test of the first-order velocity $v^r$ \eqref{eq:vr_expression_ksz} against an extended expression $v_{(2)}$ (including partial second-order terms), evaluated under an extreme stress-test profile ($m_{0,\mathrm{amp}} = 0.5$, $r_0 = 1200\,\mathrm{Mpc}$). Top panel: Comparison of the first-order velocity $v_{(1)}$ (solid blue) and $v_{(2)}$ (dashed orange, incorporating the Jacobian $d\tilde{\eta}/d\eta$ and pullback terms), showing a peak peculiar velocity of $\approx -3900\,\mathrm{km\,s^{-1}}$. Bottom panel: Velocity residual $\delta v = v_{(2)} - v_{(1)}$, bounded by $|\delta v| \lesssim 80\,\mathrm{km\,s^{-1}}$. This corresponds to a relative error of $\lesssim 2\%$ at the peak, confirming that Eq.~\eqref{eq:vr_expression_ksz} remains remarkably accurate.
}
\label{fig:app_vr_accuracy}
\end{figure}

Fig.~\ref{fig:app_vr_accuracy} demonstrates that these additional terms yield only a minor change in velocity, even for an extreme stress-test profile ($m_{0,\mathrm{amp}} = 0.5$, $r_0 = 1200\,\mathrm{Mpc}$). Their impact is even smaller within the posterior region used in the main analysis, confirming the accuracy of Eq.~\eqref{eq:vr_expression_ksz} at the required order.

\section{Effective velocity decomposition}
\label{app:velocity_decomposition}

\begin{figure}[htbp]
\centering
\includegraphics[width=0.8\textwidth]{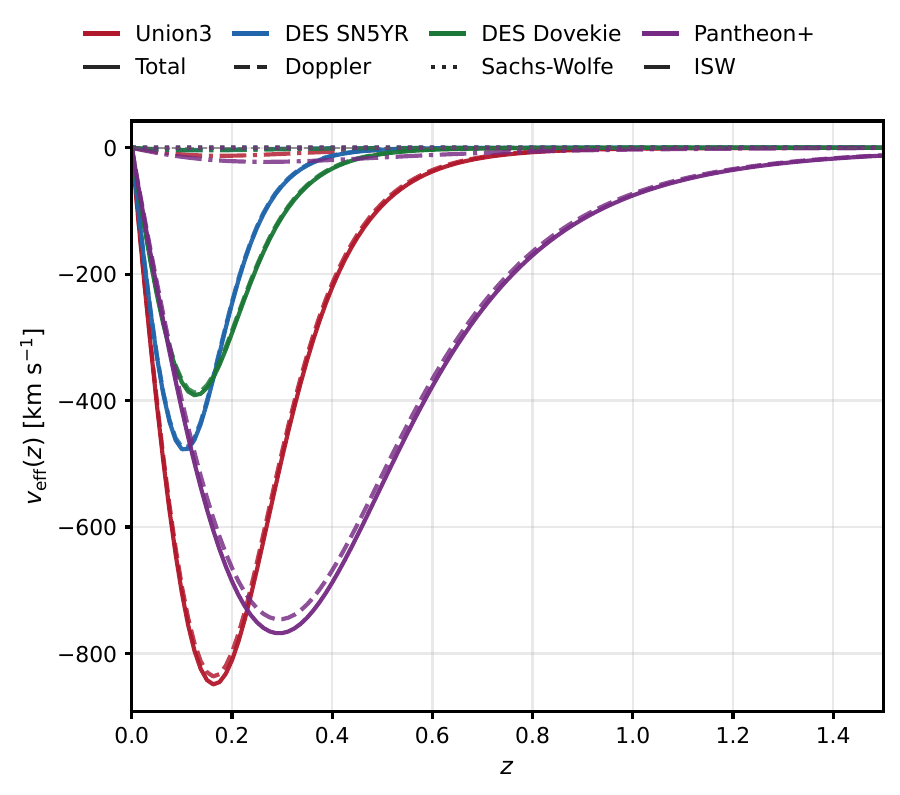}
\caption{
Redshift evolution of the effective velocity $v_{\mathrm{eff}}(z)$ for the four distance-only posterior-median profiles, decomposed into Doppler, SW, and ISW components. Colors denote datasets (Union3: red; DES-SN5YR: blue; DES Dovekie: green; Pantheon+: purple), while line styles indicate total (solid), Doppler (dashed), SW (dotted), and ISW (dash-dotted) contributions. Across all profiles, the signal is strongly dominated by the Doppler term, with the ISW contribution playing a minor secondary role and the SW term being negligible. Because the LTB kSZ power scales as $D_{\ell=3000}^{\mathrm{kSZ}} \propto \int v_{\mathrm{eff}}^2\,dz$, the higher peak amplitudes and broader redshift spans in Pantheon+ and Union3 yield substantially higher kSZ power ($5.11$ and $3.41\,\mu\mathrm{K}^2$, respectively) than those in DES SN5YR and DES Dovekie ($0.63$ and $0.51\,\mu\mathrm{K}^2$, respectively).
}
\label{fig:app_veff_decomposition}
\end{figure}

Fig.~\ref{fig:app_veff_decomposition} shows the effective velocity $v_{\rm eff}(z)$, derived from the posterior-median profiles of the four distance-only analyses. For these profiles, the signal is heavily dominated by the Doppler term. The ISW term provides only a minor secondary contribution, whereas the SW term remains virtually negligible across the plotted redshift interval.

The four velocity profiles exhibit substantial variations in both amplitude and radial extent. While their peak amplitudes depend jointly on $m_{0,\rm amp}$ and $r_0$, their width in redshift space is governed primarily by $r_0$. These trends align with the posterior medians listed in Table~\ref{tab:distance_constraints}. Specifically, a spatially broader density profile sustains a coherent velocity field over a wider redshift span, while a larger profile amplitude enhances the overall flow strength. According to Eq.~\eqref{eq:Clksz_ksz}, the LTB contribution to $D_{\ell=3000}^{\rm kSZ}$ is driven predominantly by the integral of $v_{\rm eff}^2$. Consequently, both the amplitude and redshift width of $v_{\rm eff}(z)$ directly modulate the kSZ power, accounting for the relative $D_{\ell=3000}^{\rm kSZ}$ values reported in the caption of Fig.~\ref{fig:app_veff_decomposition}.

\section{Empirical Fitting Formula for $D_{\ell=3000}^{\rm kSZ}$}
\label{app:D3000_fit}

To provide a rapid diagnostic estimate of $D_{\ell=3000}^{\rm kSZ}$ and clarify how the kSZ power constrains the density profile parameters, we construct the following empirical fitting formula
\begin{equation}
D_{3000}^{\rm fit}
\simeq
1.69\,
\exp\left[
\left(
\frac{\Delta r/r_0}{0.60}
\right)^2
\right]
\exp\left[
0.077\frac{m_{0,\rm amp}}{0.10}
\right]
\left(
\frac{m_{0,\rm amp}}{0.10}
\right)^2
\left(
\frac{r_0}{1000\,{\rm Mpc}}
\right)^{3.25}
\mu{\rm K}^2.
\label{eq:app_D3000_fit}
\end{equation}
This expression serves solely as a diagnostic surrogate for rapid $D_{\ell=3000}^{\rm kSZ}$ estimations. The background cosmological parameters are omitted from the fit, as their influence on $D_{\ell=3000}^{\rm kSZ}$ is minimal.

\begin{figure}[htbp]
\centering
\includegraphics[width=0.9\textwidth]{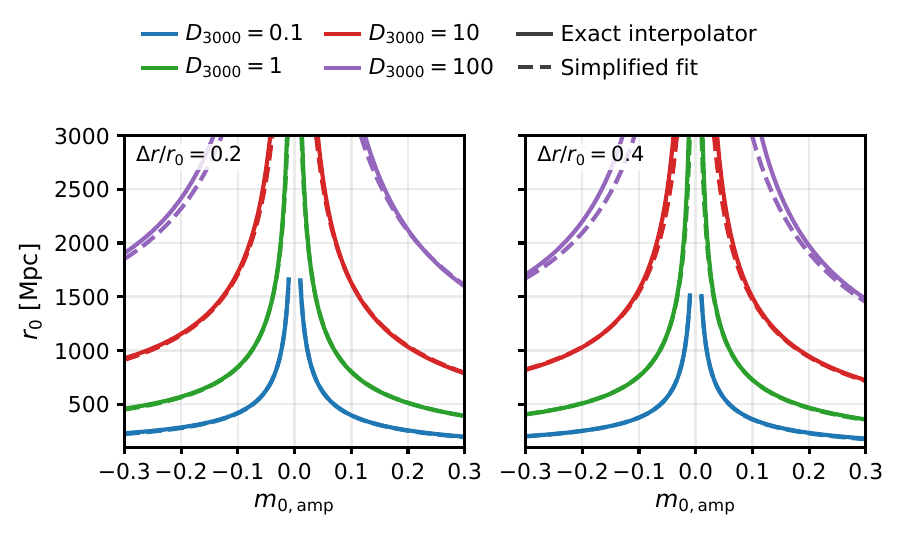}
\caption{
Fidelity test of the fitting formula in Eq.~\eqref{eq:app_D3000_fit} (dashed) against the exact numerical interpolator (solid), mapped in the $(m_{0,\mathrm{amp}}, r_0)$ plane for profile width ratios of $\Delta r/r_0 = 0.2$ (left) and $0.4$ (right). Contours mark constant kSZ power levels of $D_{\ell=3000}^{\mathrm{kSZ}} = 0.1$ (blue), $1$ (green), $10$ (red), and $100\,\mu\mathrm{K}^2$ (purple). The fitting formula agrees remarkably well with the exact solution in the relevant weak-to-moderate regime ($D_{\ell=3000}^{\mathrm{kSZ}} \le 10\,\mu\mathrm{K}^2$), diverging noticeably only in the strong-signal regime ($D_{\ell=3000}^{\mathrm{kSZ}} \sim 100\,\mu\mathrm{K}^2$).
}
\label{fig:app_D3000_fit_error}
\end{figure}

We evaluate the fidelity of Eq.~\eqref{eq:app_D3000_fit} over $m_{0,\rm amp}\in[-0.3,0.3]$ and $r_0\in[100,3000]\,{\rm Mpc}$. Fig.~\ref{fig:app_D3000_fit_error} presents representative comparisons for $\Delta r/r_0=0.2$ and $0.4$, overlaying contours from the exact numerical interpolator with those from the fitting formula.

The tight agreement among the low-amplitude contours demonstrates that Eq.~\eqref{eq:app_D3000_fit} yields reliable estimates when $D_{\ell=3000}^{\rm kSZ}$ is small. Its performance remains adequate near $D_{\ell=3000}^{\rm kSZ}=10\,\mu{\rm K}^2$. However, the growing divergence between contours at higher amplitudes reflects larger discrepancies in the strong-kSZ regime.

The leading power-law dependence on $m_{0,\rm amp}$ and $r_0$ in Eq.~\eqref{eq:app_D3000_fit} stems from the velocity scaling. From Eq.~\eqref{eq:vr_expression_ksz}, we have $v_{\rm eff}\sim r\partial_\eta B$, while Eq.~\eqref{eq:B_solution_distance} establishes $B\propto m_{0,\rm amp}$, implying $v_{\rm eff}\propto m_{0,\rm amp}r$. Approximating the redshift-weighted integral of $v_{\rm eff}^2$ as a spatial radial integration yields the scaling $D_{\ell=3000}^{\rm kSZ}\propto m_{0,\rm amp}^2 r_0^3$. The fitted exponent of $3.25$ deviates slightly from $3$ for two reasons: first, the kSZ power is integrated over redshift rather than spatial coordinates; second, $H_\parallel$, $D_M$, and $\tau$ in Eq.~\eqref{eq:Clksz_ksz} also depend on the LTB profile. Finally, the exponential factor in $m_{0,\rm amp}$ absorbs residual amplitude-dependent corrections.

\section{Profile-width robustness check with the kSZ constraint}
\label{app:width_robustness}

While our baseline analysis fixes the transition-width ratio at $\Delta r/r_0=0.4$, we test the sensitivity to this profile choice by repeating the analysis with $\Delta r/r_0=0.2$. Table~\ref{tab:app_width02_summary} summarizes the resulting parameter constraints and model-comparison metrics.

\begin{table}[htbp]
\centering
\renewcommand{\arraystretch}{1.15}
\begin{tabular}{lccccc}
\hline\hline
\shortstack{SNe~Ia compilation\\(with BAO+CMB)}
& $m_{0,\rm amp}$
& $r_0$ (Mpc)
& $\Delta{\rm AIC}_{\Lambda{\rm CDM}}$
& \shortstack{Log Bayes factor\\$\ln K_{\Lambda{\rm LTB},\Lambda{\rm CDM}}$}
& $\Delta\chi^2_{w_0w_a{\rm CDM}}$ \\
\hline
Union3
& $0.112^{+0.044}_{-0.047}$
& $600^{+150}_{-120}$
& $-7.47$
& $0.96$
& $0.65$ \\
DES SN5YR
& $0.136^{+0.045}_{-0.042}$
& $350^{+190}_{-100}$
& $-11.76$
& $3.61$
& $-0.10$ \\
DES Dovekie
& $0.095^{+0.045}_{-0.048}$
& $400^{+310}_{-150}$
& $-5.46$
& $0.80$
& $-0.59$ \\
Pantheon+
& $0.046^{+0.231}_{-0.035}$
& $480^{+1110}_{-430}$
& $-3.68$
& $-1.16$
& $-1.27$ \\
\hline
\end{tabular}%
\caption{
Same as Table~\ref{tab:ksz_d1_constraints}, but with the transition width changed from $\Delta r = 0.4 r_0$ to $\Delta r = 0.2 r_0$ as a robustness check. Parameter shifts across all SNe~Ia compilations remain within $1\sigma$ uncertainties, and the qualitative model comparison remains unchanged, validating the robustness of our baseline analysis.
}
\label{tab:app_width02_summary}
\end{table}

Reducing the fixed width ratio from $\Delta r/r_0=0.4$ to $0.2$ does not materially alter the inferred radial profile. The parameter and metric shifts are most noticeable for Union3 and Pantheon+: for Union3, $r_0$ increases from $490$ to $600\,{\rm Mpc}$ (with $m_{0,\rm amp}$ decreasing from $0.129$ to $0.112$), while for Pantheon+, $r_0$ decreases from $600$ to $480\,{\rm Mpc}$. These shifts are comparable to or smaller than the corresponding $68\%$ uncertainties. Correspondingly, model-comparison metrics for Union3 and Pantheon+ shift by up to $\sim 1.07$ in $\Delta{\rm AIC}_{\Lambda{\rm CDM}}$ and $\Delta\chi^2_{w_0w_a{\rm CDM}}$, with Pantheon+ exhibiting a $\ln K_{\Lambda{\rm LTB},\Lambda{\rm CDM}}$ shift of 0.58. For the two DES compilations, the shifts are at most $0.32$ in $\Delta{\rm AIC}_{\Lambda{\rm CDM}}$ and $\Delta\chi^2_{w_0w_a{\rm CDM}}$, and at most $0.11$ in $\ln K_{\Lambda{\rm LTB},\Lambda{\rm CDM}}$. Because the qualitative model comparison remains unchanged, this test confirms the robustness of our main conclusions to the assumed transition width, though it does not replace a continuous exploration of $\Delta r/r_0$.


\acknowledgments

Pengjie Zhang and Yukang Zhang are supported by the National Key R\&D Program of China (2023YFA1607800, 2023YFA1607801). We thank Dr. Hongbo Cai for providing code that served as a useful reference for this work.


\bibliographystyle{JHEP}
\bibliography{bibfile}

\end{document}